\documentclass[usenatbib]{mnras}
\usepackage{graphicx}
\graphicspath{{./../../figures/}}
\usepackage{float}
\usepackage{amssymb}
\usepackage{color}
\usepackage{multirow}
\usepackage{times}

\newcommand{\frii}{FR$\,$II}
\newcommand{\fri}{FR$\,$I}
\newcommand{\wphz}{$\,$W$\,$Hz$^{-1}$}

\title[Unification of LOFAR-detected AGN in Bo\"{o}tes]{Investigating the Unification of LOFAR-detected powerful AGN in the Bo\"{o}tes Field}
\author[Leah K. Morabito]{\parbox{\textwidth}{Leah K. Morabito$^{1,2}$\thanks{E-mail: leah.morabito@physics.ox.ac.uk}, 
W.~L. Williams$^{3}$, 
Kenneth J. Duncan$^{1}$, 
H.~J.~A. R\"{o}ttgering$^{1}$, 
George Miley$^{1}$, 
Aayush Saxena$^{1}$, 
Peter Barthel$^{4}$,
P.~N. Best$^{5}$,
M. Bruggen$^{6}$,
G. Brunetti$^{7}$,
K. T. Chy\.zy$^{8}$,
D. Engels$^{6}$,
M. J. Hardcastle$^{3}$,
J.~J. Harwood$^{9}$,
Matt J. Jarvis$^{2,10}$,
E.~K. Mahony$^{11,12}$,
I. Prandoni$^{7}$,
T.~W. Shimwell$^{1}$,
A. Shulevski$^{9}$,
C. Tasse$^{13,14}$ \\} \\
$^{1}$Leiden Observatory, Leiden University, P.O. Box 9513, NL-2300 RA, Leiden, The Netherlands\\
$^{2}$Astrophysics, University of Oxford, Denys Wilkinson Building, Keble Road, Oxford, OX1 3RH \\
$^{3}$Centre for Astrophysics Research, School of Physics, Astronomy and Mathematics, University of Hertfordshire, College Lane, Hatfield AL10 9AB, UK \\
$^{4}$ Kapteyn Astronomical Institute, University of Groningen, P.O.Box 800, NL - 9700AV Groningen, The Netherlands \\
$^{5}$SUPA, Institute for Astronomy, Royal Observatory, Blackford Hill, Edinburgh, EH9 3HJ, UK \\
$^{6}$Universit\"{a}t Hamburg, Hamburger Sternwarte, Gojenbergsweg 112, 21029 Hamburg, Germany \\
$^{7}$INAF- Istituto di Radioastronomia, via P. Gobetti 101, 40129 Bologna. Italy \\
$^{8}$Astronomical Observatory, Jagiellonian University, ul. Orla 171, 30-244 Krak\'ow, Poland \\
$^{9}$ASTRON, The Netherlands Institute for Radio Astronomy, Postbus 2, 7990 AA, Dwingeloo, The Netherlands \\
$^{10}$Department of Physics, University of the Western Cape, Cape Town 7535, South Africa \\
$^{11}$Sydney Institute for Astronomy, School of Physics A28, The University of Sydney, NSW 2006, Australia \\
$^{12}$ARC Centre of Excellence for All-Sky Astrophysics (CAASTRO) \\
$^{13}$GEPI, Observatoire de Paris, CNRS, Université Paris Diderot, 5 place Jules Janssen, 92190 Meudon, France \\
$^{14}$Department of Physics and Electronics, Rhodes University, PO Box 94, Grahamstown, 6140 South Africa \\
}

\definecolor{Mygrey}{gray}{0.75}

\begin{document}
\date{}
\pagerange{\pageref{firstpage}--\pageref{lastpage}} \pubyear{2014}
\maketitle

\label{firstpage}

\begin{abstract}
Low radio frequency surveys are important for testing unified models of radio-loud quasars and radio galaxies. Intrinsically similar sources that are randomly oriented on the sky will have different projected linear sizes. Measuring the projected linear sizes of these sources provides an indication of their orientation. Steep-spectrum isotropic radio emission allows for orientation-free sample selection at low radio frequencies. We use a new radio survey of the Bo\"{o}tes field at 150$\,$MHz made with the Low Frequency Array (LOFAR) to select a sample of radio sources. We identify 60 radio sources with powers $P>10^{25.5}\,$\wphz\ at 150$\,$MHz using cross-matched multi-wavelength information from the AGN and Galaxy Evolution Survey (AGES), which provides spectroscopic redshifts and photometric identification of 16 quasars and 44 radio galaxies. When considering the radio spectral slope only, we find that radio sources with steep spectra have projected linear sizes that are on average 4.4$\pm$1.4 larger than the those with flat spectra. The projected linear sizes of radio galaxies are on average 3.1$\pm$1.0 larger than those of quasars (2.0$\pm$0.3 after correcting for redshift evolution). Combining these results with three previous surveys, we find that the projected linear sizes of radio galaxies and quasars depend on redshift but not on power. The projected linear size ratio does not correlate with either parameter. The LOFAR data is consistent within the uncertainties with theoretical predictions of the correlation between the quasar fraction and linear size ratio, based on an orientation-based unification scheme. 
\end{abstract}


\section{Introduction}
\label{sec:s1}
Active galactic nuclei (AGN) produce high nuclear luminosities that cannot be explained by star formation alone. They are believed to be powered by accretion on to a central super-massive black hole. AGN can exhibit a wide variety of observational characteristics, including high luminosities in the optical and near-IR, strong emission lines from ionised gas, and high mid-IR luminosities. AGN which have strong emission lines in their spectra are classified as Type 1 or Type 2 based on whether their emission lines are broad (typically $>2000\,$km$\,$s$^{-1}$) or narrow ($<2000\,$km$\,$s$^{-1}$). Type 1 AGN show broad and narrow emission lines while Type 2 AGN show only narrow emission lines. 

Under the current unification paradigm \citep{antonucci93,up95} Type 1 and Type 2 AGN exhibit different observed characteristics due to the presence of a dusty structure or torus that will obscure the accretion disk and broad line region depending on viewing angle to the AGN. Those AGN which exhibit broad emission lines were historically associated with bright optical point sources, leading to the term `quasi-stellar object' or quasar. Type 1 AGN are oriented such that the accretion disk and broad line emission regions are seen directly without obscuration from the torus. They include radio-loud and radio-quiet quasars as well as Type 1 Seyfert galaxies. Type 2 AGN are oriented such that the broad line emission region and accretion disk are obscured by the dusty torus, and not directly visible. 

About 10 per cent of AGN exhibit extended powerful radio emission in the form of jets that reach far beyond the host galaxy and can provide an indication of orientation. Low radio frequency surveys are important for selecting these objects in an orientation-independent way, as high-frequency surveys are biased towards core-dominated flat-spectrum objects where the jets are pointed towards the observer. The radio fluxes of radio-loud AGN at low frequencies are dominated by emission from the lobes, rather than the hotspots and/or jets, thus minimising any possible orientation-based effects like Doppler boosting. These radio-loud AGN comprise two populations, believed to be powered by different accretion modes. Radio galaxies that exhibit strong emission lines in the optical regime and evidence of a torus in the mid-infrared are referred to as high-excitation radio galaxies (HERGs). HERGs can be either Type 1 or Type 2 AGN. The AGN in HERGs are thought to be powered via a geometrically thin and optically thick accretion disk \citep{shakura73}, which is believed to be fed by large central repositories of cold gas \citep[e.g.,][]{larson10}. This has led to the name `cold-mode' accretion \citep{hardcastle07}. 
 
The second type of radio-loud AGN lack the strong emission lines seen in HERGs, and termed low-excitation radio galaxies (LERGs). They also lack evidence for a dusty torus \citep[e.g.,][]{vdw10,tasse08,ogle06,whysong04}, and for a full accretion disk \citep[e.g.,][]{hardcastle06,evans06}. LERGs are thought to be powered by `hot-mode' accretion processes, where hot gas is accreted via advection-dominated accretion or radiatively inefficient accretion flows \citep[e.g.,][]{ny94,quataert01,ho08}. For these reasons the non-radio spectral energy distributions of LERGs are not expected to show strong orientation effects. \citet{bh12} showed that LERGs dominate the population of low-power ($P_{1.4\textrm{GHz}} \lesssim 10^{25}$\wphz ) sources at least in the local universe, and tend to be associated with edge-dimmed radio jets of Fanaroff-Riley Class I objects \citep[\fri ;][]{fr74}, although this is not a hard division with one-to-one mapping.   
  
The discovery of super-luminal motion in radio jets was strong evidence that some radio sources have their jets aligned close (within $\sim15\,$degrees) to the line of sight \citep{barthel89a}. Radio sources with beamed flat-spectrum cores consistent with super-luminal motion should have smaller projected sizes on average than steep-spectrum radio sources with jets oriented further from the line of sight. This scenario can be extended to HERGs, where the difference in orientation is indicated by whether or not the line of sight reveals emission from the accretion disk. Radio galaxies are those objects in which the obscuring torus is viewed edge-on and the accretion disk emission is hidden, while quasars are viewed with a direct line of sight to the accretion disk. Radio galaxies will therefore have radio jets preferentially oriented closer to the plane of the sky, while quasars will have radio jets closer to the line of sight. 

The radio orientation is clearly linked to the orientation of the torus/accretion disk, as shown by the  detection of significant optical polarisation aligned with the radio jets in nearby galaxies \citep[e.g.,][]{schmidt00,antonucci82}. The observed properties of strong line AGN (HERGs) at wavelengths other than radio are consistent with orientation schemes, mostly due to the presence of a dusty obscuring structure. For example, the presence of broad lines in polarised light of Type 1 AGN (narrow line galaxies) is powerful evidence for hidden quasars whose light is reflected outside of the obscuring structure \citep[e.g.,][]{cohen99,ogle97,antonucci84}. 

\citet{barthel89} used the 3CRR survey of radio sources at 178$\,$MHz \citep{3CRR} to study the projected linear sizes for 42 radio sources with optical identifications and spectroscopic redshifts for the range $0.5\leq z \leq 1$. Barthel found that there was a division between radio sources: those associated with  quasars were on average 2.2 times smaller than the other radio sources. The 3CRR sample is now 100 per cent spectroscopically complete, and classifications based on emission line ratios have been used to identify HERGs (which are expected to show orientation effects) and LERGs (which are not expected to show orientation effects). The high limiting flux density of the survey means that only 13 per cent have been found to be LERGs \citep{willott99}, and therefore were not a large contaminant at the redshift range used by \citet{barthel89}. 	

Follow up studies have tended to confuse the issue. A reassessment of the 3CRR sample by \citet{singal14} found systematically larger sizes of radio galaxies compared to radio-loud quasars only for redshifts above 0.5. For lower redshifts, radio galaxies were on average larger than radio-loud quasars but only when their cumulative linear sizes were above $\sim 400\,$kpc. \citet{singal13a} also studied the linear sizes of a 98 per cent spectroscopically complete sample selected at 408$\,$MHz \citep{blr99}, finding that only at redshifts larger than 1 were the linear sizes of radio galaxies systematically larger than those of quasars. The authors made no attempt to remove LERGs from the sample, but argued that they would have to be a large part of the sample to change the results. However these studies still only cover a small part of the power$-$redshift ($P-z$) diagram, and it is important to collect more information to investigate this further. 

A further motivation for larger studies comes from \citet{dipompeo13}. The authors use Monte Carlo modelling to show that intrinsic size distributions and the intrinsic angle of division between radio galaxies and quasars, $\theta_c$, can influence the results. While they do not consider in their models that $\theta_c$ should correlate directly with the linear size ratio, they estimate that for $\theta_c=45^{\circ}$ \citep[][found $\theta_c=44.4^{\circ}$]{barthel89} several hundred sources are necessary for a significant difference in the cumulative measured sizes. Their comparison to other data sets that seem to contradict orientation-only unification schemes \citep[][and an unpublished study]{singal13a} does not consider observational biases (e.g., classification of LERGs/HERGs) that might be present in the data. Here we try to overcome these biases by using a new low-frequency selected sample, and by combining and considering all available data for a broader test of orientation-based unification schemes.

In this study we use a new low frequency radio survey of the Bo\"{o}tes field \citep{williams16} with the Low Frequency Array \citep[LOFAR;][]{vh13} to investigate the unification of radio sources in two ways. First, we look at the difference in projected linear sizes of flat- and steep-spectrum sources, without any classification from other wavelengths. Second, we investigate the difference in projected linear sizes of cold-mode accretion sources by splitting our sample into \emph{quasars} and \emph{radio galaxies} using available multi-wavelength data including spectroscopic redshifts and quasar classifications from the AGN and Galaxy Evolution Survey \citep[AGES;][]{kochanek12}. It is difficult to measure properly the sizes of edge-dimmed \fri\ sources, and we make a power cut at $P_{150\textrm{MHz}} >10^{25.5}$\wphz\ to exclude these sources as well as star-forming galaxies at lower powers (Saxena et al., in preparation) as much as possible. This means our sample is likely to be dominated by HERGs, although it is difficult to determine this based on broad-band photometry alone \citep[Ch. 5 of][]{janssen17}. 

The new LOFAR catalogue contains over 6,000 radio sources to an rms depth of $\sim120-150\,\mu$Jy beam$^{-1}$ at 150$\,$MHz. While previous samples used to test orientation-based unification probed high luminosity sources over large areas of the sky, the deep LOFAR data probes a large number of fainter sources over a smaller area. This adds a substantial number of sources to the P$-z$ diagram at lower powers compared to previous samples, especially for higher redshifts. 
 
Section~\ref{sec:s2} first describes the LOFAR survey, the multi-wavelength data, and the selection of quasars and radio galaxies. Results from the LOFAR survey are presented in Section \ref{sec:s3}. The LOFAR results combined with two previous samples are presented in Section~\ref{sec:s4}. Discussion and conclusions follow in Sections \ref{sec:s5} and \ref{sec:s6}.  Throughout the paper we assume a $\Lambda$CDM concordance cosmology with $H_0=67.8$ km$\,$s$^{-1}\,$Mpc$^{-1}$, $\Omega_{\textrm{m}}=0.308$, and $\Omega_{\Lambda}=0.692$, consistent with \citet{planckcosmo}. Spectral index is defined as $\alpha$ with flux density $S\propto\nu^{\alpha}$. Throughout the paper, all `linear sizes' referred to are projected linear sizes.

\section{The Bo\"{o}tes Field Data}
\label{sec:s2}
In this section, we present the construction of our radio galaxy and quasar samples. We start by describing the new LOFAR Bo\"{o}tes survey, the cross-matching of radio and pre-existing multi-wavelength data, and how we select our sample. Finally we discuss possible biases in our sample selection.

\subsection{LOFAR Bo\"{o}tes Survey}
\label{sec:s21}
The catalogue from \citet{williams16} contains a total of 6,267 radio sources within 19 deg$^2$.  The rms varies over the field of view, and sources with peak fluxes exceeding a threshold of 5$\sigma$ above the local rms were included in the final catalogue. The rms is less than 120$\,\mu$Jy beam$^{-1}$ at the centre and more than 50 per cent of the field of view has rms noise less than 180$\,\mu$Jy beam$^{-1}$. The average rms noise at the edges of the optical coverage is approximately 150$\,\mu$Jy beam$^{-1}$. The resolution of the LOFAR image is 5.6$\times$7.4 arcsec, with average positional accuracy of $\sim$0.4 arcsec. 
 
The sources in the catalogue are divided into classes based on radio morphology. Here we consider only single sources or extended sources with a radio core (VClass 1/11) and double sources with no obvious radio core (VClass 2/21). These will be mostly \frii , \fri , or single component sources. The other morphological classifications only make up 0.1 per cent of the catalogue. These sources have either diffuse or complex morphologies, and not expected to be the radio sources in which we are interested. 

\subsection{Adding in Multi-wavelength Data}
\label{sec:s22}
To carry out this study, we require two pieces of information: spectroscopic redshifts, and the ability to split our LOFAR-detected radio-loud AGN into quasars and radio galaxies. The spectroscopic redshifts are necessary to calculate precise linear sizes of radio sources, and avoid propagating large and uncertain errors due to photometric redshifts. The quasar/radio galaxy classifications are necessary to divide the sources into the desired samples. Ideally, we would target all LOFAR-detected sources to acquire both spectroscopic redshift and enough information to distinguish spectroscopically between radio galaxies (narrow-line AGN) and quasars (broad- and narrow-line AGN). A dedicated survey with the William Herschel Telescope Enhanced Area Velocity Explorer (WEAVE) which goes online in 2018 \citep[WEAVE-LOFAR][]{weave} will provide spectroscopic redshifts and possibly emission line ratios for LOFAR-detected sources within the next few years. As these data do not exist yet, we use available multi-wavelength information. 

The NOAO Deep Wide-Field Survey \citep[NDWFS;][]{jd99} covers 9 deg square of the Bo\"{o}tes field with deep optical to near-infrared photometric data ($B_W$, $R$, $I$, $J$, $K$). Ancillary data at longer wavelengths covers the near to mid-infrared (IRAC 3.6, 4.5, 5.8, and 8.0$\mu$m, and MIPS 24$\mu$m). A multi-wavelength catalogue of these data, based on \citet{brown07,brown08} was used for the optical identification of radio sources, described in Williams et al. (submitted). Within the NDWFS sky coverage, there are 3,894 LOFAR-detected sources; of these, 76 per cent, or 2,971 sources, have an optical counterpart. 

The AGN and Galaxy Evolution Survey \citep[AGES;][]{kochanek12} is based on the NDWFS data, as well as complementary ultraviolet, radio, and X-ray data. The NDWFS contains more than 2 million optical sources, and  \citet[][hereafter K12]{kochanek12} aimed to provide a statistically robust sample of normal galaxies via random sparse sampling, and a complete sample of targeted AGN candidates (selected from the combination of multi-wavelength photometric data). These samples were targeted for spectroscopic redshift measurements using the Hectospec instrument on the MMT. From 8,977 AGN candidates, spectra were taken for 7,102 and redshifts obtained for 4,764 (after excluding Galactic stars, which made up 9 per cent of successful redshifts). Measured spectroscopic redshifts are therefore available for $\sim 53$ per cent of AGN candidates in AGES.  In total there are spectroscopic redshifts for 18,163 galaxies (to $I=$20 mag) and 4,764 AGN candidates (to $I=$22.5 mag). There are 1,106 LOFAR-detected sources with optical counterparts which have spectroscopic redshifts in AGES.

\subsection{Selecting the Final Samples}
\label{sec:s23}

The way we select our final sample is as follows:
\begin{enumerate}
\item Select all LOFAR-detected sources with optical counterparts.
\item Select only optical counterparts with spectroscopic redshifts from AGES.
\item Use the K12 classification codes to identify quasars.
\item Those sources not identified as quasars are considered to be radio galaxies.
\item Check that all radio galaxies satisfy the mid-IR 24$\,\mu$ criteria, which indicates the presence of a dusty torus (and are therefore not likely to be LERGs).
\item Make a radio power cut of $P\geq10^{25.5}\,$W$\,$Hz$^{-1}$ to exclude \fri\ and star-forming galaxies.
\end{enumerate}

For all sources with spectroscopic information, K12 provide classification codes. There are five AGN classifications which rely on different photometric bands of the multi-wavelength data (described in detail in K12). AGN can be selected based on compact, bright optical morphology, near-IR colour selections \citep[based on][]{stern05}, bright mid-IR luminosities, the presence of X-ray point sources, and radio detections at 1.4$\,$GHz. A small fraction of sources that were initially part of the normal galaxies sample are also identified as AGN in the catalogue. We use these AGN classifications to identify quasars as those objects satisfying the optical and/or near-IR colour criteria (in line with above), and the rest of our sample is therefore defined as radio galaxies. We check that all of the radio galaxies in our final sample satisfy the mid-IR 24$\,\mu$m criteria (as above), which indicates the presence of a dusty torus (and these sources are therefore not likely to be LERGs). 

For this study, we are only interested in powerful \frii\ sources. The last step in constructing our sample is to make a power cut at $10^{25.5}$\wphz\ to remove the lower power sources which are either star forming galaxies or more likely to have \fri\ morphology (since \fri\ sources are dimmer at the edges, their linear sizes are not well defined). This cut is suggested as appropriate by previous data and theoretical models (Saxena et al., submitted). To test whether we can divide the sample into higher power bins, we construct radio luminosity functions (RLFs) following the grid-based method in \citet{rigby15}. A comparison of the RLFs for our sample with model RLFs from \citet{rigby11} shows that the observed RLF for the LOFAR sample with $P_{150} \geq 10^{25.5}$\wphz\  has similar behaviour to the model RLF, i.e. the trends in space density are the same, suggesting that the sample is representative of the entire population. Introducing higher power cuts in the LOFAR sample changes the observed RLF, as the space density of radio sources in the LOFAR sample drops off sharply above $z=2$ with higher power cuts. We therefore do not consider higher power cuts in this analysis. The final sample comprises 44 radio galaxies and 16 quasars, and their properties are listed in Table Tab.~\ref{tab:t0}. 

Not all sources are resolved by this low-frequency survey, as identified by \citet{williams16}. Twenty-one sources are unambiguously detected as having edge-brightened lobes typical of \frii\ sources, 25 sources do not have individually resolved components but are clearly extended, and 14 sources are considered to be truly unresolved (point sources). Many of the unresolved sources have extended, edge-bright morphology that suggests they would be identified as \frii\ sources if imaged at higher resolution. Eight of the unresolved sources are quasars. In the case of the unresolved sources, we use the deconvolved Major axis as the largest angular size. The power cut makes it highly likely they are \frii\ type sources, and we leave them in the sample. 

\begin{table*}
\centering 
 \vspace{-12pt} \caption{\label{tab:t0} Final sample parameters. LOFAR ID is Source\_id in the LOFAR catalogue. RA and Dec is from the AGES catalogue. Spectral index is given for those sources present in the WSRT catalogue. The classification is: U -- unresolved; UE -- unresolved but extended morphology; FRII -- Fanaroff-Riley II.}
 \vspace{-6pt}\begin{tabular}{lcccccrrcr} 
 \hline 
\hline 
\multicolumn{9}{c}{Quasars} \\
\hline 
LOFAR ID & $z$ & Optical RA & Optical Dec & LAS [arcsec] & LLS [kpc] & Flux density [mJy] & Power [W Hz$^{-1}$] & $\alpha$ & Class \\[-3pt] 
\hline 
909 & 1.876 &  219.54611 &   34.08321 &   2.4 &  20.4 & $ 494.62\pm 1.9$ &  $ 4.47\pm0.02\times 10^{27}$ & -0.53 & UE \\
1105 & 3.244 &  219.37666 &   34.96291 &   2.6 &  20.2 & $   9.89\pm 0.3$ &  $ 2.28\pm0.08\times 10^{26}$ & -0.50 & U \\
1197 & 1.485 &  219.30660 &   35.09860 &   6.9 &  59.6 & $  92.15\pm 0.4$ &  $ 5.43\pm0.03\times 10^{26}$ & -0.15 & UE \\
2136 & 2.187 &  218.64273 &   35.16937 &   2.1 &  17.9 & $  17.73\pm 0.2$ &  $ 2.10\pm0.02\times 10^{26}$ &  0.62 & U \\
2988 & 2.865 &  218.12744 &   34.58065 &   2.3 &  18.6 & $   8.94\pm 0.3$ &  $ 1.68\pm0.05\times 10^{26}$ & -1.27 & U \\
3414 & 2.049 &  217.86592 &   34.62810 &   3.0 &  25.6 & $  15.18\pm 0.3$ &  $ 1.60\pm0.03\times 10^{26}$ & -0.57 & UE \\
4287 & 2.352 &  217.31340 &   34.63901 &   2.6 &  22.0 & $   2.48\pm 0.2$ &   $ 3.34\pm0.3\times 10^{25}$ & -1.05 & U \\
4302 & 2.207 &  217.29259 &   35.49635 &   2.3 &  19.4 & $  12.78\pm 0.3$ &  $ 1.54\pm0.03\times 10^{26}$ &  0.29 & U \\
4424 & 1.036 &  217.21750 &   34.86668 &   2.6 &  21.6 & $  11.25\pm 0.3$ &  $ 3.33\pm0.09\times 10^{25}$ & -0.36 & U \\
4474 & 0.747 &  217.17730 &   35.72420 &  26.9 & 202.4 & $  43.74\pm 0.7$ &   $ 6.74\pm0.1\times 10^{25}$ &     & UE \\
4590 & 0.413 &  217.10619 &   34.92972 &  28.4 & 160.3 & $  74.39\pm 0.7$ &  $ 3.40\pm0.03\times 10^{25}$ & -0.80 & UE/FRII \\
4769 & 2.156 &  216.96770 &   35.50896 &   2.5 &  21.6 & $   5.25\pm 0.2$ &   $ 6.07\pm0.3\times 10^{25}$ & -0.59 & U \\
5349 & 1.255 &  216.43303 &   33.92596 &  25.8 & 221.0 & $1651.05\pm12.8$ &  $ 7.08\pm0.05\times 10^{27}$ & -0.73 & UE \\
5426 & 1.207 &  216.35101 &   34.16002 &   3.4 &  28.9 & $  15.63\pm 0.4$ &   $ 6.21\pm0.1\times 10^{25}$ &     & U \\
5442 & 2.787 &  216.31882 &   34.87961 &  11.1 &  89.5 & $ 188.49\pm 1.0$ &  $ 3.38\pm0.02\times 10^{27}$ &     & UE \\
5512 & 1.601 &  216.23466 &   35.47827 &  30.7 & 266.5 & $  65.32\pm 0.5$ &  $ 4.42\pm0.03\times 10^{26}$ &     & FRII \\
\hline 
\hline 
\multicolumn{9}{c}{Radio Galaxies} \\
\hline 
LOFAR ID & $z$ & Optical RA & Optical Dec & LAS [arcsec] & LLS [kpc] & Flux density [mJy] & Power [W Hz$^{-1}$] & $\alpha$ & Class \\[-3pt] 
\hline 
825 & 0.439 &  219.61089 &   33.84379 &  79.4 & 464.6 & $  73.65\pm 0.5$ &  $ 3.82\pm0.03\times 10^{25}$ & -0.69 & FRII \\
1093 & 0.739 &  219.38063 &   34.83463 &  22.6 & 169.8 & $  87.49\pm 0.5$ & $ 1.32\pm0.01\times 10^{26}$ & -0.53 & UE \\
1148 & 0.579 &  219.34889 &   35.12636 &  23.8 & 161.3 & $ 107.98\pm 0.6$ &  $ 9.91\pm0.06\times 10^{25}$ &     & UE \\
1269 & 2.648 &  219.21390 &   33.65285 &   4.8 &  39.5 & $   3.79\pm 0.2$ &   $ 6.24\pm0.4\times 10^{25}$ & -0.40 & UE \\
1536 & 0.846 &  219.01352 &   33.73015 &  45.9 & 361.1 & $ 174.87\pm 0.6$ &  $ 3.46\pm0.01\times 10^{26}$ & -1.23 & FRII \\
1578 & 1.132 &  218.99646 &   34.75127 &  18.3 & 154.2 & $  43.44\pm 0.4$ &  $ 1.53\pm0.01\times 10^{26}$ & -0.68 & UE \\
1639 & 2.117 &  218.94856 &   33.88606 &   5.0 &  42.8 & $   9.18\pm 0.3$ &  $ 1.03\pm0.03\times 10^{26}$ & -0.24 & UE \\
1739 & 1.290 &  218.86841 &   33.32543 &  10.9 &  93.5 & $  37.41\pm 0.5$ &  $ 1.69\pm0.02\times 10^{26}$ & -0.17 & FRII \\
1758 & 0.686 &  218.87160 &   34.57289 &  30.8 & 224.3 & $ 147.89\pm 1.2$ &  $ 1.92\pm0.01\times 10^{26}$ & -0.50 & UE \\
1823 & 1.120 &  218.81210 &   33.36397 &   1.7 &  14.4 & $  25.75\pm 0.4$ &   $ 8.86\pm0.1\times 10^{25}$ & -0.78 & U \\
2145 & 2.464 &  218.63110 &   35.01800 &  13.0 & 107.5 & $   9.42\pm 0.3$ &  $ 1.37\pm0.04\times 10^{26}$ & -0.73 & FRII \\
2406 & 1.674 &  218.47987 &   34.15944 &   8.8 &  76.7 & $  11.94\pm 0.3$ &   $ 8.77\pm0.2\times 10^{25}$ & -0.56 & UE \\
2493 & 0.494 &  218.42067 &   33.73990 &  35.5 & 221.5 & $  61.22\pm 0.4$ &  $ 4.05\pm0.03\times 10^{25}$ & -0.62 & FRII \\
2642 & 0.310 &  218.31844 &   34.85692 & 124.3 & 583.7 & $1515.39\pm 1.5$ & $ 3.82\pm0.004\times 10^{26}$ & -0.92 & FRII \\
2643 & 1.609 &  218.31567 &   33.48305 &  36.7 & 318.8 & $  32.38\pm 0.4$ &  $ 2.21\pm0.03\times 10^{26}$ & -0.63 & FRII \\
2705 & 0.497 &  218.29622 &   33.97461 &  29.7 & 186.1 & $  70.08\pm 0.5$ &  $ 4.70\pm0.03\times 10^{25}$ & -0.45 & UE \\
2712 & 2.409 &  218.29308 &   33.76791 &   2.4 &  19.7 & $   5.12\pm 0.2$ &   $ 7.18\pm0.3\times 10^{25}$ & -0.73 & U \\
2720 & 0.488 &  218.29130 &   35.25504 &  33.1 & 205.6 & $ 186.03\pm 0.9$ & $ 1.20\pm0.006\times 10^{26}$ & -0.76 & UE/FRII \\
2958 & 0.479 &  218.14023 &   33.71775 &  34.3 & 210.8 & $  80.30\pm 0.7$ &  $ 4.99\pm0.04\times 10^{25}$ & -0.72 & UE \\
2974 & 2.244 &  218.13596 &   33.98435 &   2.3 &  19.4 & $   3.59\pm 0.3$ &   $ 4.45\pm0.3\times 10^{25}$ & -0.95 & U \\
3278 & 1.982 &  217.95490 &   33.16585 &   2.9 &  24.6 & $   5.50\pm 0.4$ &   $ 5.48\pm0.4\times 10^{25}$ & -0.39 & U \\
3348 & 1.775 &  217.91176 &   35.59362 &   1.6 &  13.9 & $   5.01\pm 0.3$ &   $ 4.10\pm0.2\times 10^{25}$ & -0.96 & UE \\
3552 & 0.863 &  217.79130 &   33.88386 &   7.4 &  58.2 & $ 136.50\pm 1.6$ &  $ 2.81\pm0.03\times 10^{26}$ & -0.67 & UE \\
3614 & 0.726 &  217.74939 &   35.36936 &  90.3 & 673.6 & $  36.70\pm 0.4$ &  $ 5.34\pm0.05\times 10^{25}$ & -0.60 & FRII \\
3631 & 2.324 &  217.74316 &   33.51855 &   3.2 &  26.5 & $   7.56\pm 0.2$ &   $ 9.96\pm0.3\times 10^{25}$ & -0.98 & UE \\
3671 & 0.510 &  217.71825 &   33.22300 &  88.2 & 560.3 & $ 654.94\pm 1.5$ &  $ 4.63\pm0.01\times 10^{26}$ & -0.73 & FRII \\
3918 & 1.826 &  217.54826 &   35.00576 &  23.3 & 201.9 & $  59.28\pm 0.7$ &  $ 5.10\pm0.06\times 10^{26}$ & -0.60 & FRII \\
3925 & 0.778 &  217.55066 &   33.24461 &  35.5 & 271.8 & $1014.18\pm 2.9$ & $ 1.70\pm0.005\times 10^{27}$ & -0.76 & FRII \\
3940 & 3.248 &  217.53720 &   34.78715 &   4.1 &  31.3 & $ 116.00\pm 1.1$ &  $ 2.68\pm0.02\times 10^{27}$ & -0.63 & U \\
4058 & 0.628 &  217.45322 &   35.29644 &  42.6 & 299.3 & $  53.56\pm 0.4$ &  $ 5.80\pm0.05\times 10^{25}$ & -0.63 & FRII \\
4081 & 1.125 &  217.42775 &   33.94859 & 101.8 & 857.9 & $ 104.62\pm 0.7$ &  $ 3.63\pm0.03\times 10^{26}$ & -0.85 & FRII \\
4137 & 0.656 &  217.41853 &   32.96680 &  64.1 & 458.7 & $  51.56\pm 0.5$ &  $ 6.11\pm0.05\times 10^{25}$ & -0.58 & FRII \\
4294 & 0.320 &  217.29977 &   33.44387 & 153.2 & 735.2 & $ 302.35\pm 0.7$ &  $ 8.15\pm0.02\times 10^{25}$ & -0.95 & FRII \\
4544 & 2.023 &  217.14195 &   34.34181 &   2.7 &  23.4 & $   6.44\pm 0.3$ &   $ 6.66\pm0.3\times 10^{25}$ &     & U \\
4577 & 0.487 &  217.12979 &   33.15258 &  38.7 & 239.9 & $  68.24\pm 0.4$ &  $ 4.39\pm0.03\times 10^{25}$ & -0.85 & FRII \\
4708 & 0.868 &  217.02147 &   35.22464 &   5.7 &  45.1 & $  25.58\pm 0.3$ &  $ 5.33\pm0.06\times 10^{25}$ & -0.68 & UE/FRII \\
4831 & 1.239 &  216.93529 &   33.64130 &   4.2 &  35.5 & $  22.30\pm 0.4$ &   $ 9.33\pm0.2\times 10^{25}$ & -0.16 & UE \\
5037 & 0.809 &  216.75853 &   33.20626 &  29.3 & 227.2 & $  34.10\pm 0.4$ &  $ 6.17\pm0.07\times 10^{25}$ & -0.58 & FRII \\
5040 & 0.570 &  216.75694 &   33.09038 &  39.2 & 263.5 & $  59.51\pm 0.8$ &  $ 5.29\pm0.07\times 10^{25}$ &     & UE \\
5142 & 1.351 &  216.64010 &   33.53679 &  29.9 & 257.8 & $  35.62\pm 0.6$ &  $ 1.76\pm0.03\times 10^{26}$ & -0.61 & FRII \\
5171 & 0.259 &  216.58580 &   34.67144 & 118.1 & 488.5 & $ 352.47\pm 1.3$ &  $ 6.13\pm0.02\times 10^{25}$ & -0.63 & FRII \\
5337 & 0.328 &  216.42273 &   34.97419 &  99.0 & 483.0 & $1020.23\pm 5.4$ &  $ 2.89\pm0.02\times 10^{26}$ & -0.61 & FRII \\
5374 & 0.688 &  216.38220 &   35.56160 &  33.1 & 241.8 & $  42.54\pm 0.5$ &  $ 5.55\pm0.06\times 10^{25}$ &     & UE \\
5411 & 1.202 &  216.38695 &   33.02372 &   5.8 &  49.2 & $  25.96\pm 0.6$ &  $ 1.02\pm0.02\times 10^{26}$ &     & UE \\
\hline 
\end{tabular} 
\end{table*}

\subsection{Investigating Possible Biases in the Sample Selection}
\label{sec:s24}

We identify quasars (and thus radio galaxies) based on the photometric information in AGES. The complicated AGES sample selection, and the fact that it changed several times over the duration of the survey, means it is not straightforward to estimate the completeness of our final sample. Nevertheless we investigate several aspects to identify any biases that might exist in our final sample. 

First we select all sources in AGES, regardless of LOFAR detections, and identify sources with identical criteria to selecting the radio galaxy and quasar samples. This allows us to see if the selection criteria preferentially favour successful spectroscopic redshifts for one sample relative to the other. We find similar spectroscopic completeness of both samples as a function of both (i) $Ks$-band magnitude  and (ii) $I$-band magnitude. This implies that we are not biased towards one sample over the other because of apparent brightness. However, the magnitude cut itself may introduce a bias in the quasar fraction if quasars are intrinsically brighter in the $I$-band than radio galaxies. Powerful radio galaxies are often found in massive, red galaxies which will be harder to detect at redshifts $\gtrsim$1, which could artificially increase the quasar fraction at higher redshifts. 

\begin{figure*}
\includegraphics[width=\columnwidth]{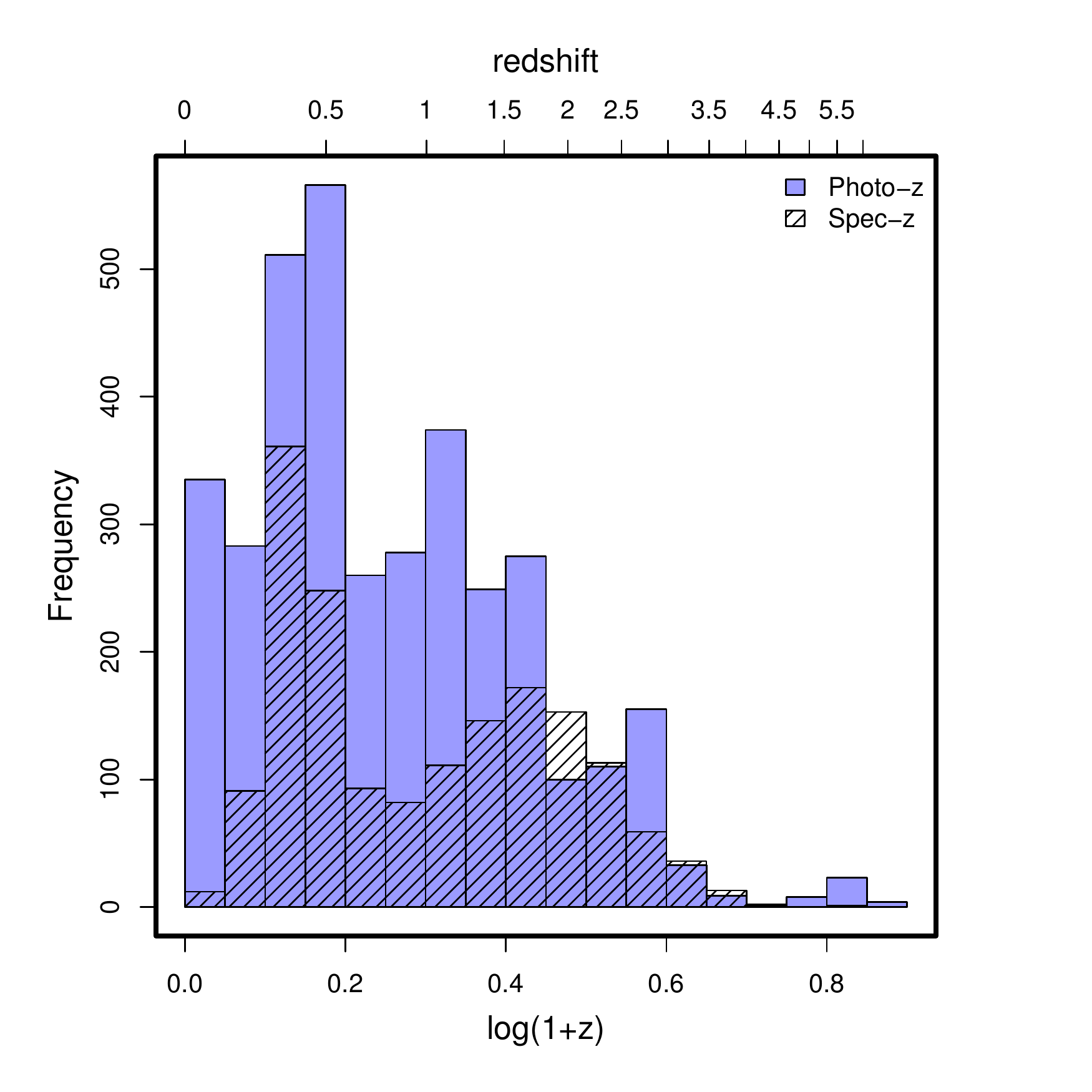}
\includegraphics[width=0.9\columnwidth]{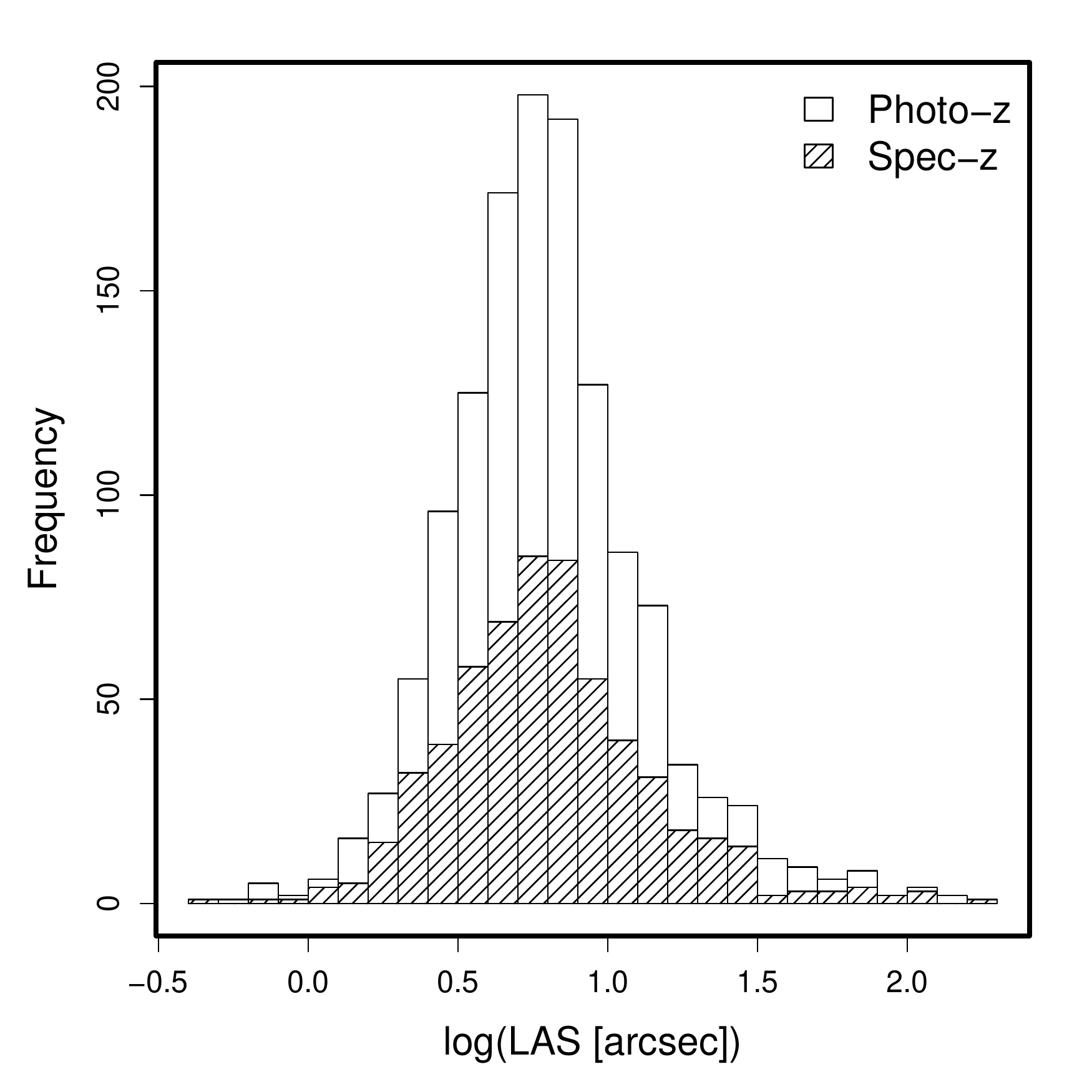}
\caption[Comparison of redshift and radio angular size for LOFAR-detected AGN in AGES]{\label{fig:f1} Comparison of redshift and radio angular size distributions for AGN from AGES which have $I\leq 22.5$ and LOFAR-detected radio emission. \emph{Left:} Distribution of redshifts for sources with photometric redshifts (unfilled histogram) and spectroscopic redshifts (hatched histogram). \emph{Right:} Distribution of radio largest angular size (LAS) for sources with photometric redshifts (unfilled histogram) and spectroscopic redshifts (hatched histogram).  }
\end{figure*}

Next we compare the distributions of spectroscopic redshifts for the two samples. The distributions have similar log-normal behaviour, as expected from a flux-limited survey, with fewer objects from both samples found at higher redshifts. For the redshift range $0.4\lesssim z \lesssim 1$ we find a slight dip in the quasar redshift distribution. To investigate this, we compare the spectroscopic redshift distribution for the quasar sample to photometric redshifts for the same sample. The photometric redshifts are from Duncan et al. (in preparation), and are combined from estimates produced using \textsc{eazy} \citep{brammer08} and three different sets of templates. These template sets are \emph{(a)} the default set of \textsc{eazy} templates, \emph{(b)} the SWIRE template library from \citet{polletta07}, and \emph{(c)} the `Atlas of Galaxy SEDs' from \citet{brown14}. The final photometric redshifts are a Hierarchical Bayesian combination of the results of the individual photometric redshift catalogues. The photometric redshift distribution shows the expected log-normal behaviour and does not show the dip that is seen in the spectroscopic redshift distribution. The comparison indicates that the spectroscopic completeness falls from about 48 per cent to about 40 per cent  in this redshift range. The loss of sources in this redshift range is only about one tenth of the entire spectroscopic quasar sample, and this is not likely to strongly bias our general results. The comparison of spectroscopic and photometric redshift distributions for the quasar sample are show in the left panel of Fig.~\ref{fig:f1}. 

 We attribute this slight loss of secure quasar redshifts to the frequency coverage of the Hectospec instrument, which is 3200\AA -- 9200\AA . At this redshift range, the H$\alpha$ spectral line no longer falls in the Hectospec wavelength coverage, and the fraction of sources with multiple robust line detections results will reduce for all source types. For bright sources with continuum detection in the spectra, the quasars have a relatively featureless continuum whereas the radio galaxies typically have massive old stellar populations with absorption features, potentially leading to the lower success rate for quasars.
 
We select our sample at low radio frequencies to avoid biases from Doppler boosting of certain components of the radio sources. Doppler boosting would tend to inflate the observed power of a radio source, which could result in a larger sample size, and a higher probability of selecting sources which are not HERGs. However, at low frequencies the total flux density is dominated by the lobes, which are not Doppler boosted. Typical core-dominance values for radio sources with the same type of power as the LOFAR sample \citep[e.g.,][]{baldi13} imply that any Doppler boosted components would provide a negligible contribution to our overall sample selection. We are confident that our results are unbiased by not performing any Doppler corrections. 

Finally, the largest angular sizes (LAS) of sources in the sample were measured as the diameter of the smallest circle enclosing the source at a level of 5$\sigma$ above the local rms. The centre of the circle may or may not correspond to the optical source. The size of a small fraction ($\sim5$ per cent) of bent sources may be underestimated since we do not measure along the jets, but this will have a negligible impact on the overall trends. Our results rely on using redshift and LAS to calculate the proper linear size of radio sources. We therefore compare the distributions of the LAS for sources with spectroscopic and photometric redshifts (see right panel of Fig.~\ref{fig:f1}). These distributions are quite similar, indicating that by using only the spectroscopic redshifts were are not missing populations of systematically smaller or larger objects. We therefore do not expect our results to change if the sample were expanded. 

\section{Results}
\label{sec:s3}
In this section we present the results from the final LOFAR sample. First we investigate the projected linear sizes of flat and steep-spectrum radio sources, without any further knowledge of the type of host galaxy. We then investigate whether the projected linear sizes of quasars and radio galaxies, as identified from the multi-wavelength data, are different. This equates to a test of unification through orientation of HERGs. In all cases we use only the 60 sources with $P_{150} \geq 10^{25.5}$\wphz . 

\subsection{Flat vs. steep-spectrum Sources}
\label{sec:s31}
\begin{figure*}
\begin{center}
\includegraphics[width=0.45\textwidth]{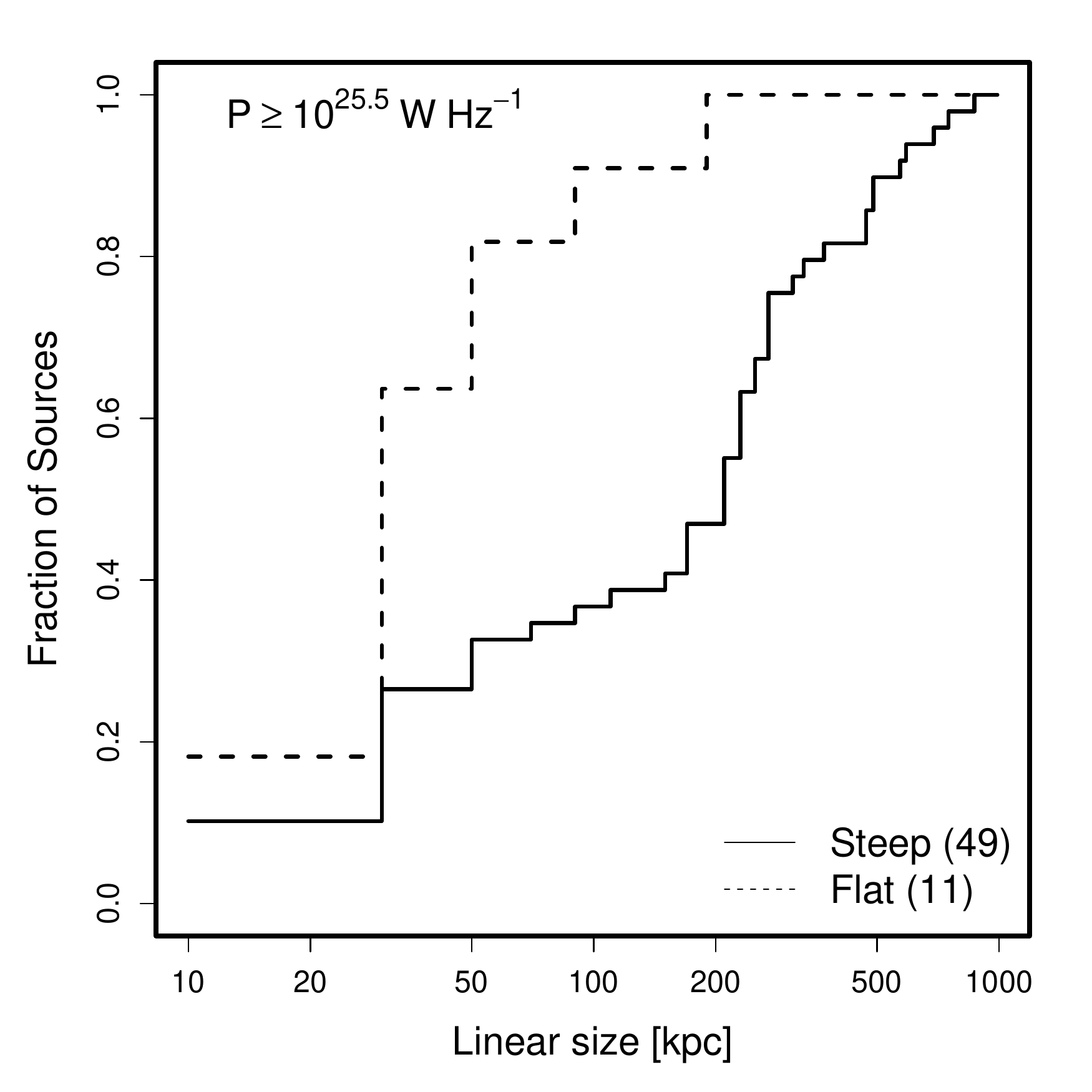}
\includegraphics[width=0.45\textwidth]{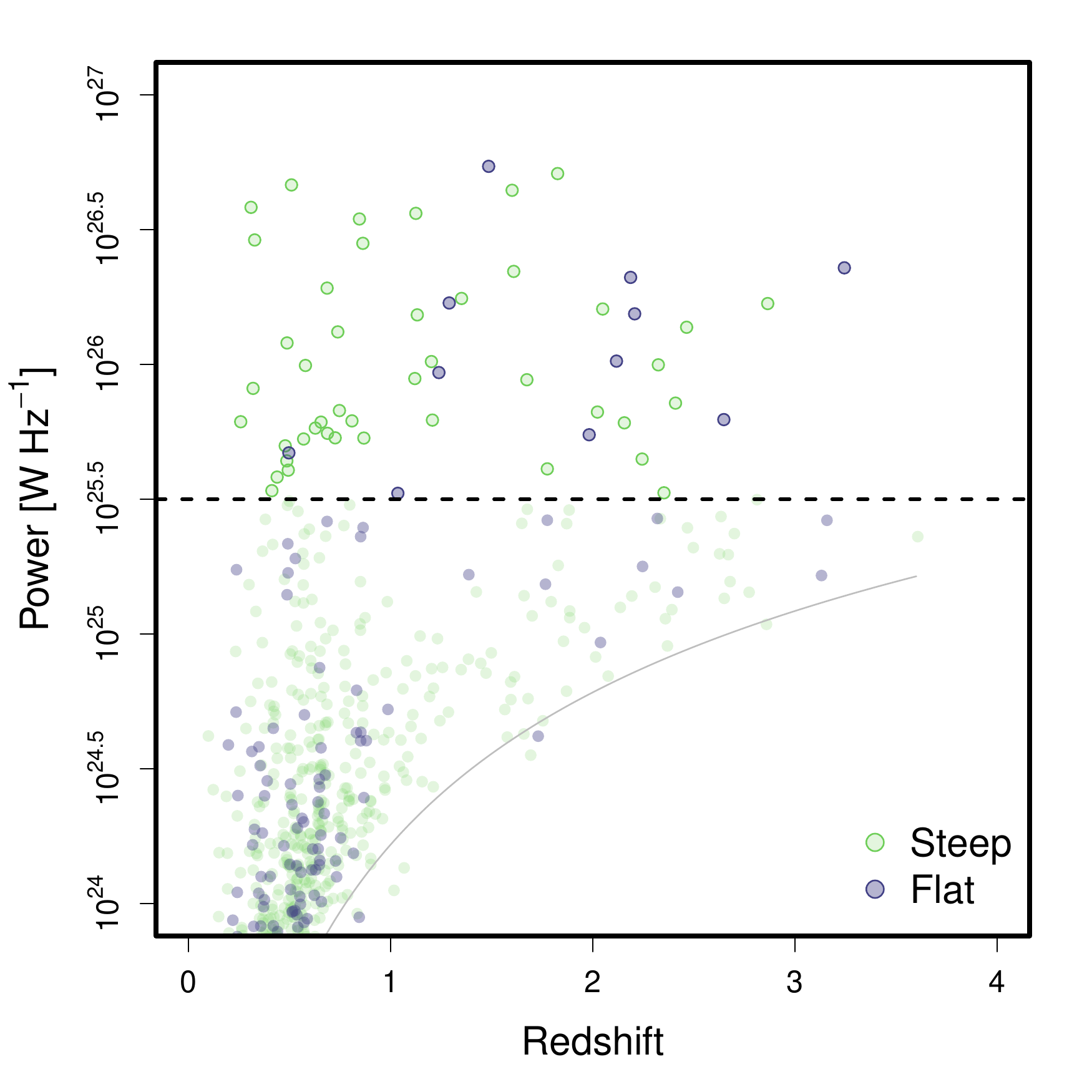}
\caption{\label{fig:f2}\emph{Left:} Cumulative linear sizes for flat-spectrum and steep-spectrum radio sources are shown by the solid and dashed lines, respectively. The number of sources in each sample is indicated in parentheses in the legends. \emph{Right:} The $P-z$ diagram. The horizontal dashed line indicates the power cut of $10^{25.5}$\wphz . The light gray line shows the flux limit for the survey. }
\end{center}
\end{figure*}

First we test whether there is a systematic difference in projected linear sizes of flat- and steep-spectrum sources. Flat-spectrum sources are likely to be beamed objects with their radio jets close to the line of sight. We use the complementary 1.4$\,$GHz data from a deep Westerbork survey of the Bo\"{o}tes field \citep{devries02}, to calculate the spectral index between 150$\,$MHz and 1.4$\,$GHz. There are 11 flat-spectrum objects with $\alpha > -0.5$ (18 per cent). The rest of the sources are considered to be steep-spectrum, and also in the case of no Westerbork detection. Fig.~\ref{fig:f2} shows the cumulative projected linear sizes of the flat and steep-spectrum objects and the distribution of sources in the $P-z$ plane. The average redshifts of the two types of sources are $z=1.23$ for the steep-spectrum sources and $z=1.81$ for the flat-spectrum sources. The mean projected linear sizes of the steep- and flat-spectrum sources are $222.42\pm29.81\,$kpc and $50.97\pm15.12\,$kpc, respectively. The projected linear sizes of steep-spectrum sources are therefore on average 4.4$\pm$1.4 times larger than those of the flat-spectrum sources. 

For a population of randomly oriented sources we can calculate the angle (between 0 and 90 degrees) which would define the division between the two populations. Qualitatively this equates to the probability that a source is oriented such that its jets are within a cone angle of $\phi$ from the line of sight, $P(\theta < \phi)=1-$cos$\phi$. The relative numbers of flat- and steep-spectrum sources then directly give the angle that divides the two populations. We calculate this angle to be $\theta_c=35.3_{- 5.9 }^{+ 5.2 }$ degrees. The average angles for the steep- and flat-spectrum populations are 65.9$^{\circ}$ and $24.7^{\circ}$, respectively. Alternatively, the opening angle can be found from the linear size ratio directly. In this case, we find $\theta_c=16.3^{+8.2}_{-4.1}$ degrees, with average angles of 61.3$^{\circ}$ and 11.5$^{\circ}$ for the steep- and flat-spectrum populations, respectively.

\subsection{Quasars vs. Radio Galaxies}
\label{sec:s32}
\begin{figure*}
\begin{center}
\includegraphics[width=0.45\textwidth]{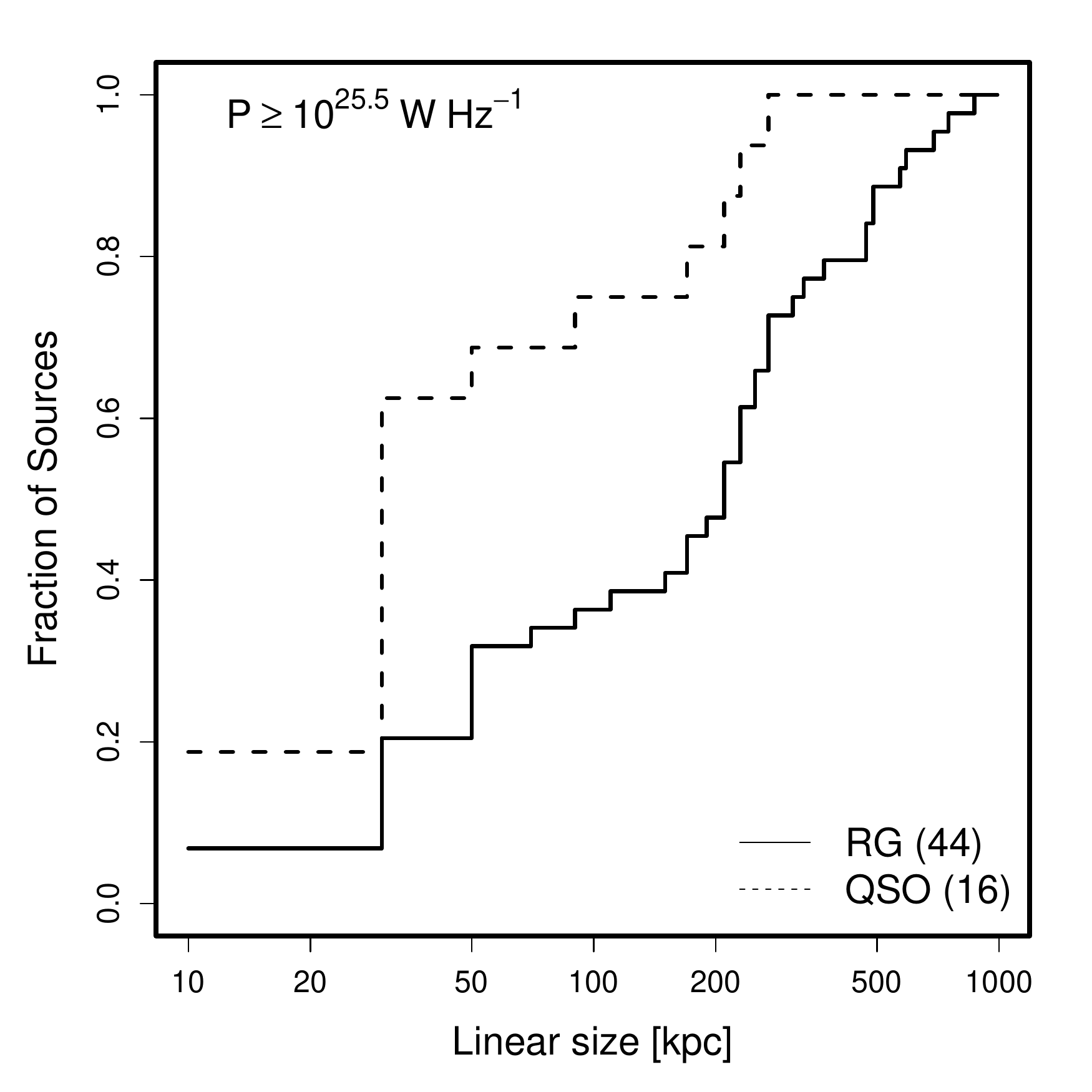}
\includegraphics[width=0.45\textwidth]{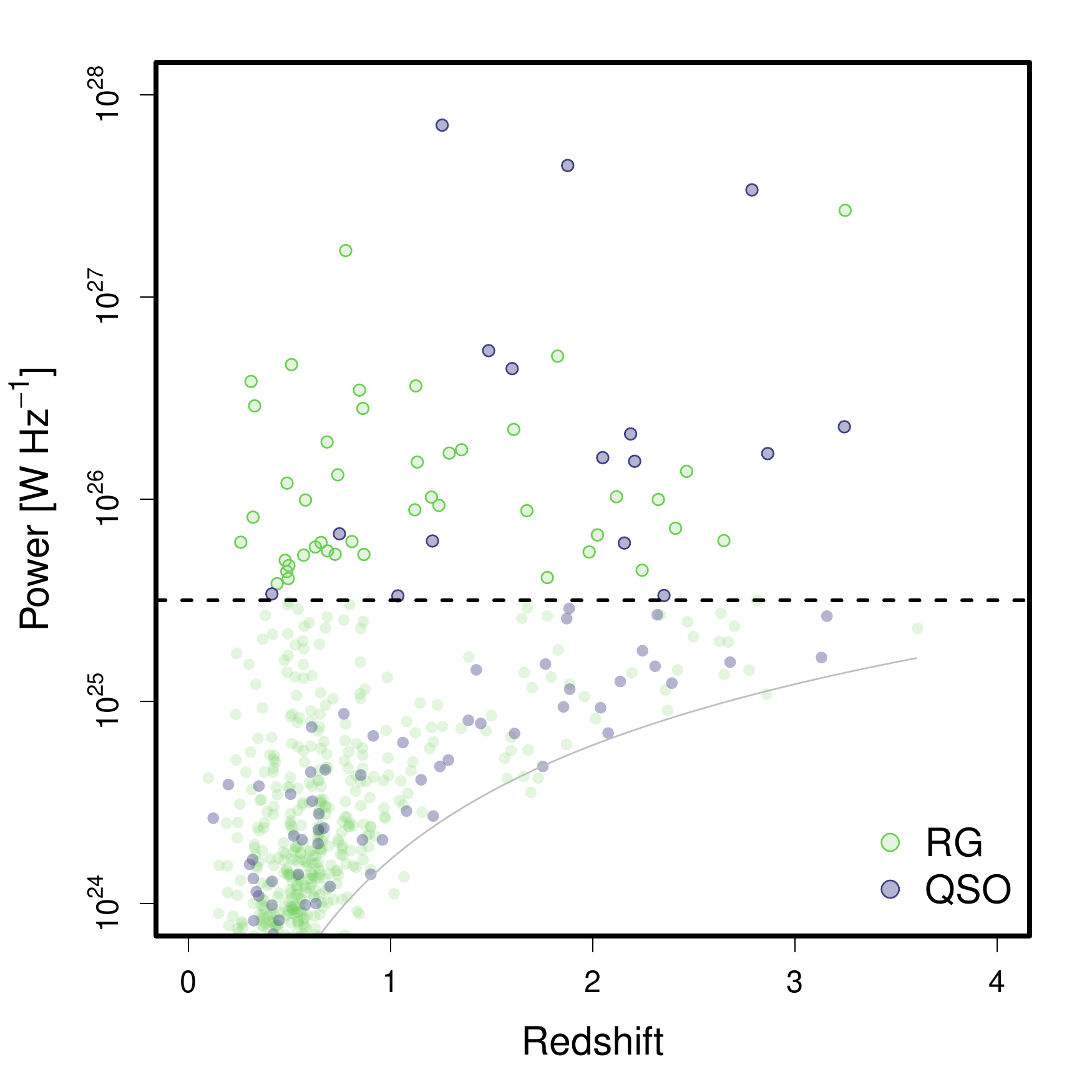}
\caption{\label{fig:f3}\emph{Left:} Cumulative linear sizes for radio galaxies (RG) and quasars (QSO) are shown by the solid and dashed lines, respectively. The number of sources in each sample is indicated in parentheses in the legends. \emph{Right:} The $P-z$ diagram. The horizontal dashed line indicates the power cut of $10^{25.5}$\wphz . The light gray line shows the flux limit for the survey. }
\end{center}
\end{figure*}

Next we test whether there is a systematic difference in the projected linear sizes of radio galaxies and  quasars. We identify the quasars and radio galaxies using the AGES criteria as described in Section~\ref{sec:s23}. The final sample has 44 radio galaxies and 16 quasars above a power cut of $10^{25.5}$\wphz . The cumulative linear sizes and distribution in the $P-z$ plane are shown in Fig.~\ref{fig:f3}. The radio galaxies have a mean projected linear size of $ 232.81 \pm 32.35\,$kpc, while the quasars have a mean projected linear size of $ 75.97 \pm 21.46\,$kpc. This corresponds to the projected linear sizes of radio galaxies being on average 3.1$\pm$1.0 times larger than those of quasars.  The relative numbers of the sources indicate the angle of division between the two classes is $\theta_c=42.8 _{- 6.0 }^{+ 5.4 }$ degrees. The average angles for the radio galaxy and quasar samples are $68.5^{\circ}$ and $29.9^{\circ}$, respectively. Using the size ratio to calculate the division yields $23.7^{+13.3}_{-6.1}$ degrees, with average angles for the radio galaxy and quasar samples of 62.7$^{\circ}$ and 16.7$^{\circ}$, respectively. 

The average redshifts of the two samples are $z=1.16$ and $z=1.84$ for radio galaxies and quasars, respectively. A detailed discussion of redshift evolution is deferred to \S~\ref{sec:s42}, but here we also report the linear size ratio after correcting the average linear sizes for the two samples to the same average redshift. After this correction the linear size ratio becomes $2.0\pm0.3$. The division angle is thus $39.2^{+9.2}_{-6.1}$ degrees, which yields average angles for the radio galaxy and quasar samples of 67.2$^{\circ}$ and 27.4$^{\circ}$, respectively.

\section{Combining LOFAR data with other Samples}
\label{sec:s4}
The LOFAR sample of sources comes from an area-limited survey, and is missing sources of higher power as these are much rarer. To account for potential radio-power dependencies in the investigation of orientation-based unification theories of radio galaxies and quasars requires the addition of data in other parts of the $P-z$ plane. In the future the LOFAR Two-metre Sky Survey \citep[LoTSS][]{lotss} will provide coverage of the entire Northern Sky, but here  we use previous radio surveys to demonstrate what this additional information will provide. 

In this section we briefly describe other low-frequency radio surveys, which cover wider areas to higher flux limits than the LOFAR survey. These surveys are all highly spectroscopically complete, have subsets of quasars already identified, and enough sources for simple binning in redshift. We inspect the cumulative linear sizes of radio galaxies and quasars in these surveys individually, and finally combine the data from our LOFAR survey and several of the previous surveys and investigate trends in projected linear size with power and redshift. Finally, we compare our observational results with the predictions of an orientation-only unification scenario. 

\subsection{Other Low-Frequency Samples}
\label{sec:s41}
The general properties of the previous surveys we use are summarised in Tab.~\ref{tab:t2}. Briefly, these surveys are: 
\begin{itemize}
\item \textbf{3CRR}  (178$\,$MHz), the revised revised 3rd Cambridge catalogue of radio sources. This is the original survey used by \citet{barthel89} to show the larger cumulative sizes of radio galaxies when compared to quasars, for the redshift range $0.5<z<1$. Since then, much more detailed optical information has become available, and the sample is now 100 per cent spectroscopically complete. In addition, near infrared spectroscopy allowed for the identification of HERGs and LERGs, using the equivalent width of $[$O$\;\textrm{III}]$ and the ratio of $[$O$\;\textrm{II}]$/$[$O$\;\textrm{III}]$  \citep{willott99,grimes04}. 
\item \textbf{7CRS}, the 7th Cambridge Redshift Survey. This comprises three separate samples (7CI,7CII,7CIII) which have been combined. These samples have similar flux limits of 0.51, 0.48, and 0.50$\,$Jy for 7CI, 7CII, and 7CIII respectively. The data for 7CI and 7CII come from \citet{grimes04}, while the 7CIII data are taken from \citet{lacy99}. Here we use the compilation from \citet{ker12}, which has LERG and quasar identifications.
\item \textbf{MRC} (408$\,$MHz), the Molonglo Radio Catalogue. This survey is complete to 1$\,$Jy with a complete quasar subset identified \citep{MRCpart3,MRCpart4}, which we use to identify quasars in the sample. This is the only sample observed at $\nu>178\,$MHz, and the only sample which does not provide HERG/LERG identification. The inclusion of LERGs may add radio sources with random orientation (i.e., not preferentially in the plane of the sky) to the sample. Since LERGs either lack or have a truncated accretion disk, they are less likely to be classified as quasars and therefore will slightly decrease the average size of radio galaxies. 
\end{itemize}

\begin{figure}
\includegraphics[width=0.48\textwidth]{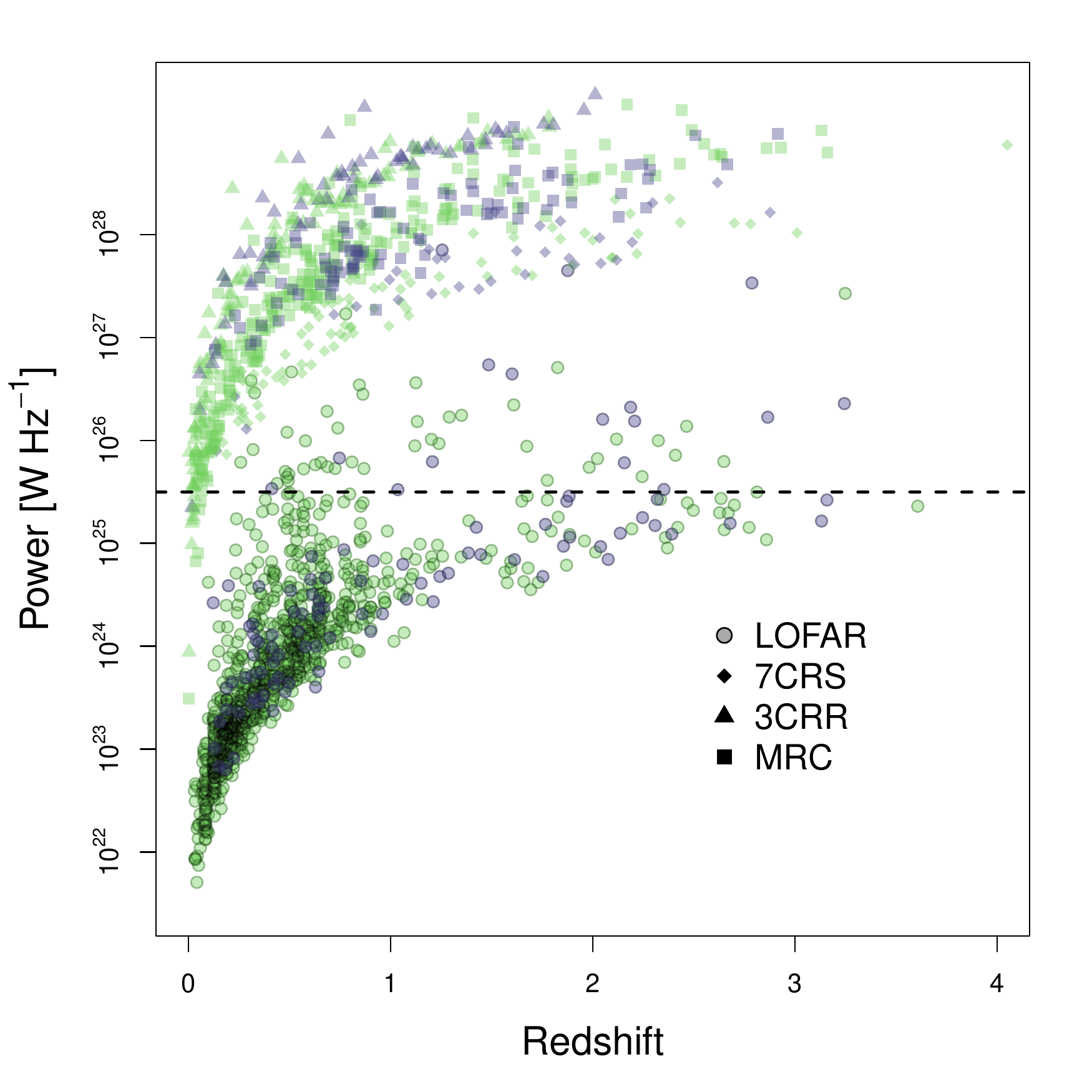}
\caption{\label{fig:f4} Power vs. redshift for the three samples described in this section plus the LOFAR sample. Only sources with spectroscopic redshifts are used, and above powers of $P_{150}>$10$^{25.5}\,$W$\,$Hz$^{-1}$ at 150$\,$MHz. Spectral index information was used to convert measured power to $P_{150}$ for the MRC and 3CRR samples. Green points represent radio galaxies and purple points represent quasars. }
\end{figure}

Overall, these surveys are highly spectroscopically complete (92 - 100 per cent, see Tab.~\ref{tab:t2}) and allow for the selection of quasars as a separate class from radio galaxies. With the exception of MRC the surveys also provide HERG/LERG classification and we are able to remove LERGs from the samples. These surveys occupy space in the $P-z$ plane which has a range of powers at very low redshifts and a smaller range of higher powers at all redshifts, see Fig.~\ref{fig:f4}. 

\begin{table*}
\caption{\label{tab:t2} A list of the previous surveys used in this paper. We use the data compiled by \citet{ker12} for 7CRS, although the original references are listed in the table. The number of sources given in the table is the number that have been identified as either HERGs (whenever possible) or quasars, and excludes LERGs, \fri\ sources, and those without spectroscopic information. }
\begin{tabular}{lcccccc}
\hline 
Survey & Frequency & Flux & \# Sources & Spectroscopic & HERG/LERG & References \\
 &	& limit	&	& completeness & separation &  \\ \hline
3CRR & 178$\,$MHz & 10.9$\,$Jy & 170 & 100\% & Yes & \cite{willott99}\footnote{Available at \\ \url{http://astroherzberg.org/people/chris-willott/research/3crr/}}, \cite{grimes04} \\
7CRS & 151$\,$MHz & $\sim$0.50$\,$Jy & 114 & 92\% & Yes & \cite{willott03,grimes04,lacy99} \\
MRC & 408$\,$MHz & 1$\,$Jy & 352 & 98\% & No & \cite{MRCpart2,MRCpart3,MRCpart4} \\ \hline
\end{tabular}
\end{table*}

\subsection{Results}
\label{sec:s42}
\begin{figure*}
\includegraphics[width=0.44\textwidth]{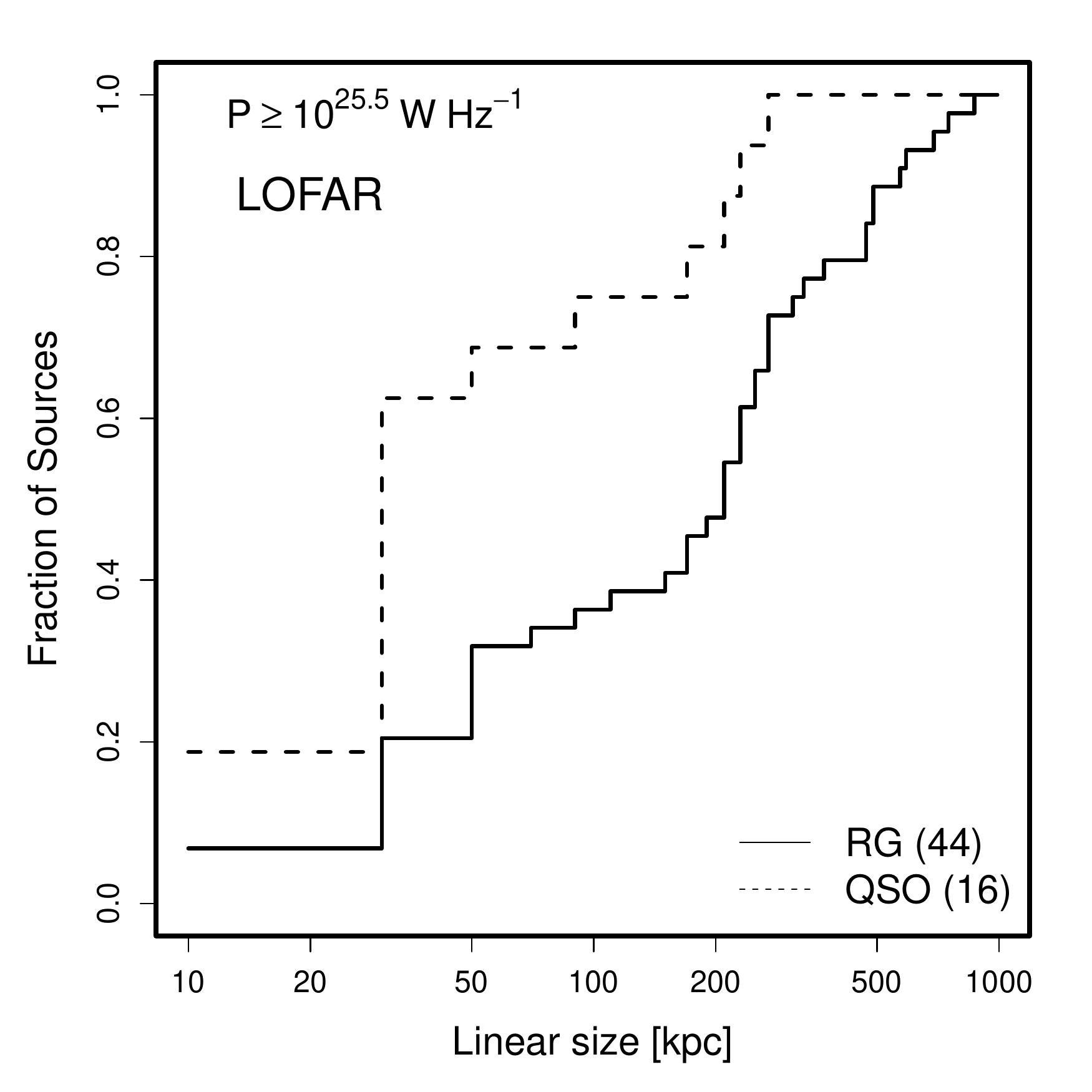}
\includegraphics[width=0.44\textwidth]{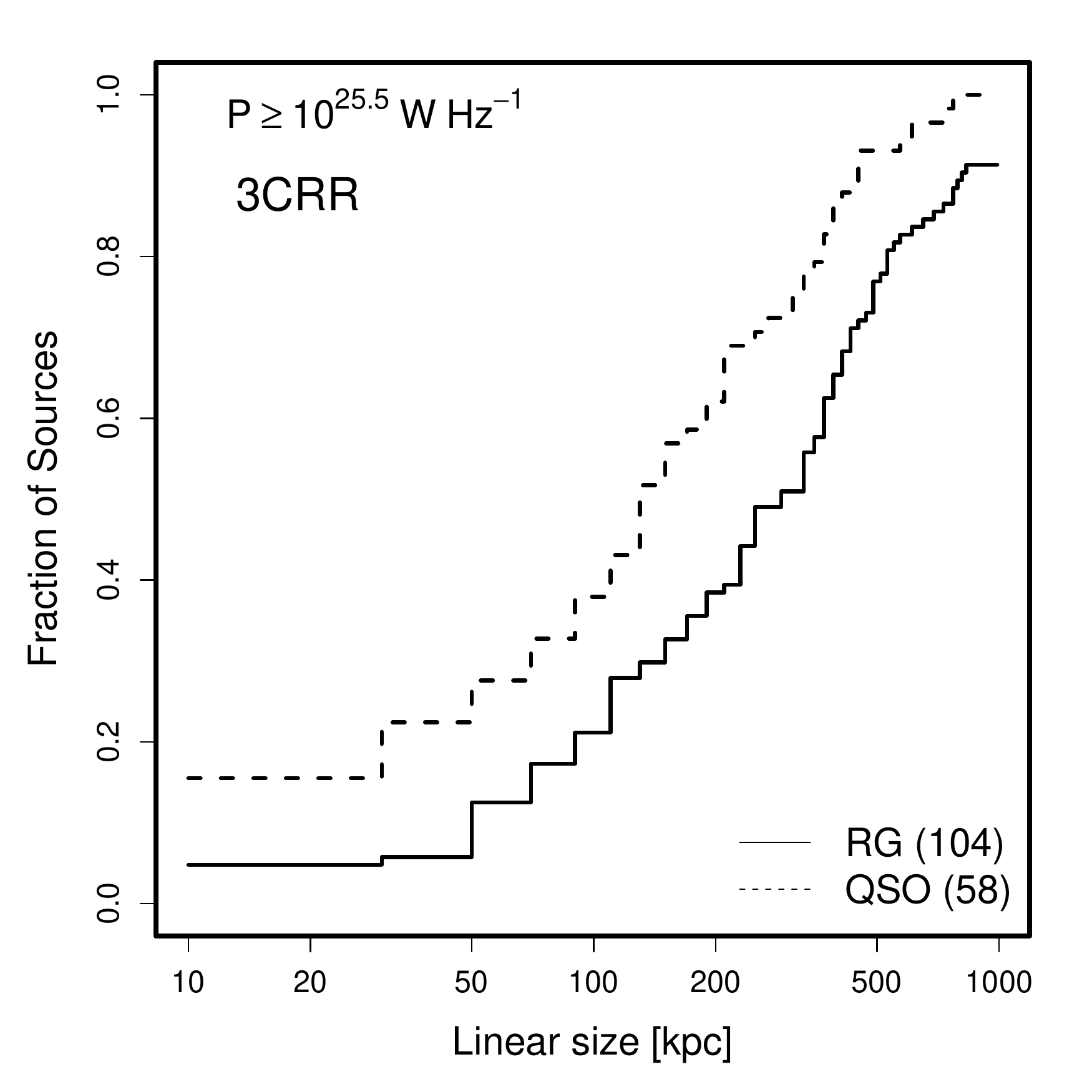} \\
\includegraphics[width=0.44\textwidth]{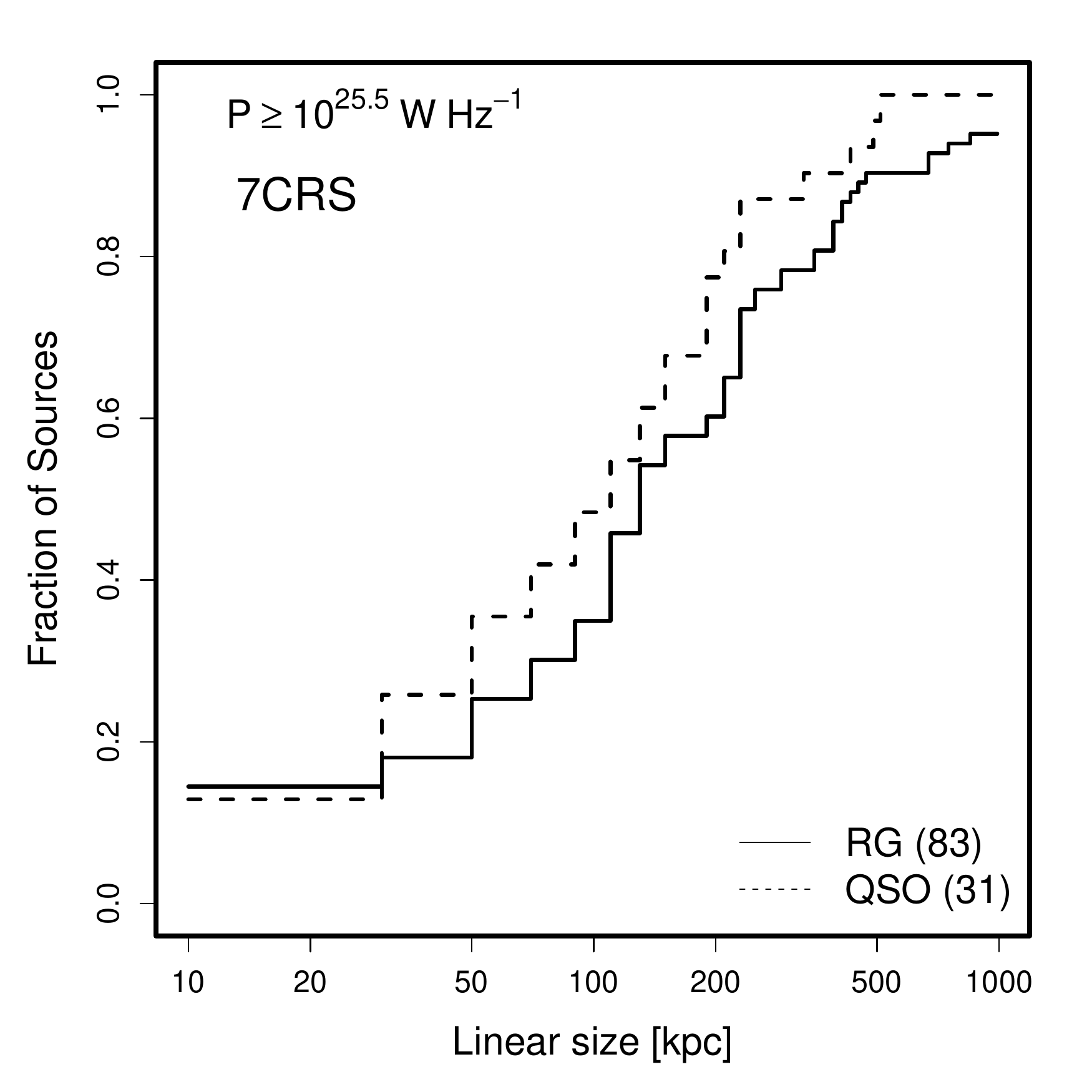}
\includegraphics[width=0.44\textwidth]{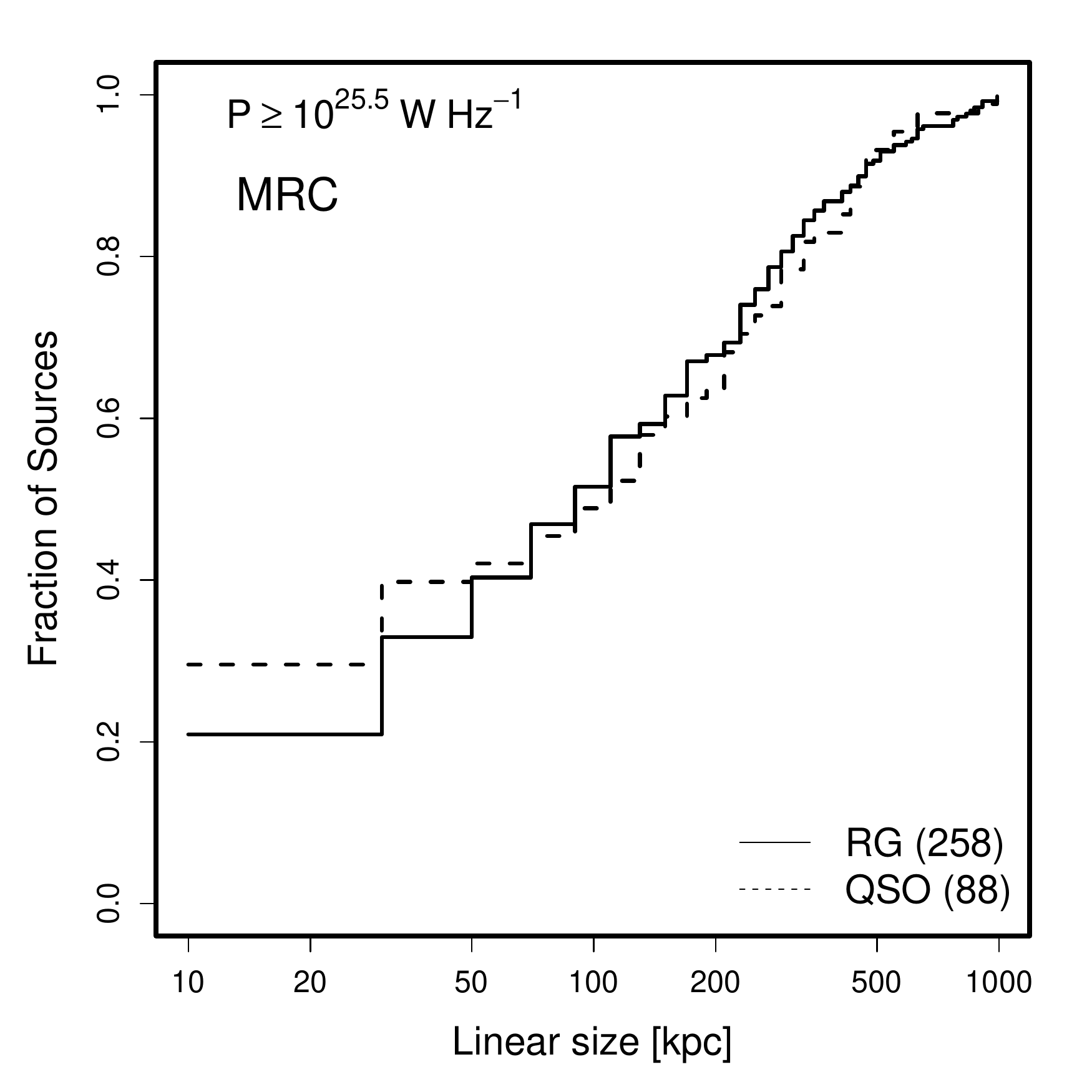}
\caption{\label{fig:f5} The cumulative linear sizes for all samples.}
\end{figure*}

We repeat the exercise of calculating the cumulative linear sizes for radio galaxies and quasars for each sample separately. These are shown in Fig.~\ref{fig:f5} for all samples, with the cumulative linear sizes from the LOFAR sample for comparison. For all samples we applied the same power cut of $P>10^{25.5}$\wphz . The LOFAR and 3CRR samples all show cumulative linear sizes of quasars that are smaller than the cumulative linear sizes of radio galaxies. The same is true for 7CRS for almost the entire range of linear sizes. The MRC sample shows that for almost the entire range of linear sizes, quasars and radio galaxies have the same cumulative linear sizes.

As each sample is considered complete by itself we treat them separately, dividing each sample into low and high redshift bins. We determined the dividing redshift for each sample such that about half of the sample lies in each redshift bin, but without letting the number of quasars drop below 5 in a bin. The dividing redshifts used are: $z=1.5$ (LOFAR), $z=0.5$ (3CRR), $z=1.5$ (7CRS), and $z=1$ (MRC).  In each bin, we count the number of radio galaxies and quasars, and measure the mean values of power, redshift, and projected linear sizes for each type of source. We calculate the ratio of projected linear sizes of radio galaxies to quasars. The total number of sources in a bin and the number of quasars in that bin are used to calculate the quasar fraction. The uncertainties were determined via standard error propagation methods, except for the case of the quasar fractions for the LOFAR sample, where we used small-number counting uncertainties following the prescription of \citet{gehrels86}. We assume that the error in spectroscopic redshift is negligible compared to the other measurement errors, and set this to zero. 

\begin{figure*}
\includegraphics[width=0.45\textwidth]{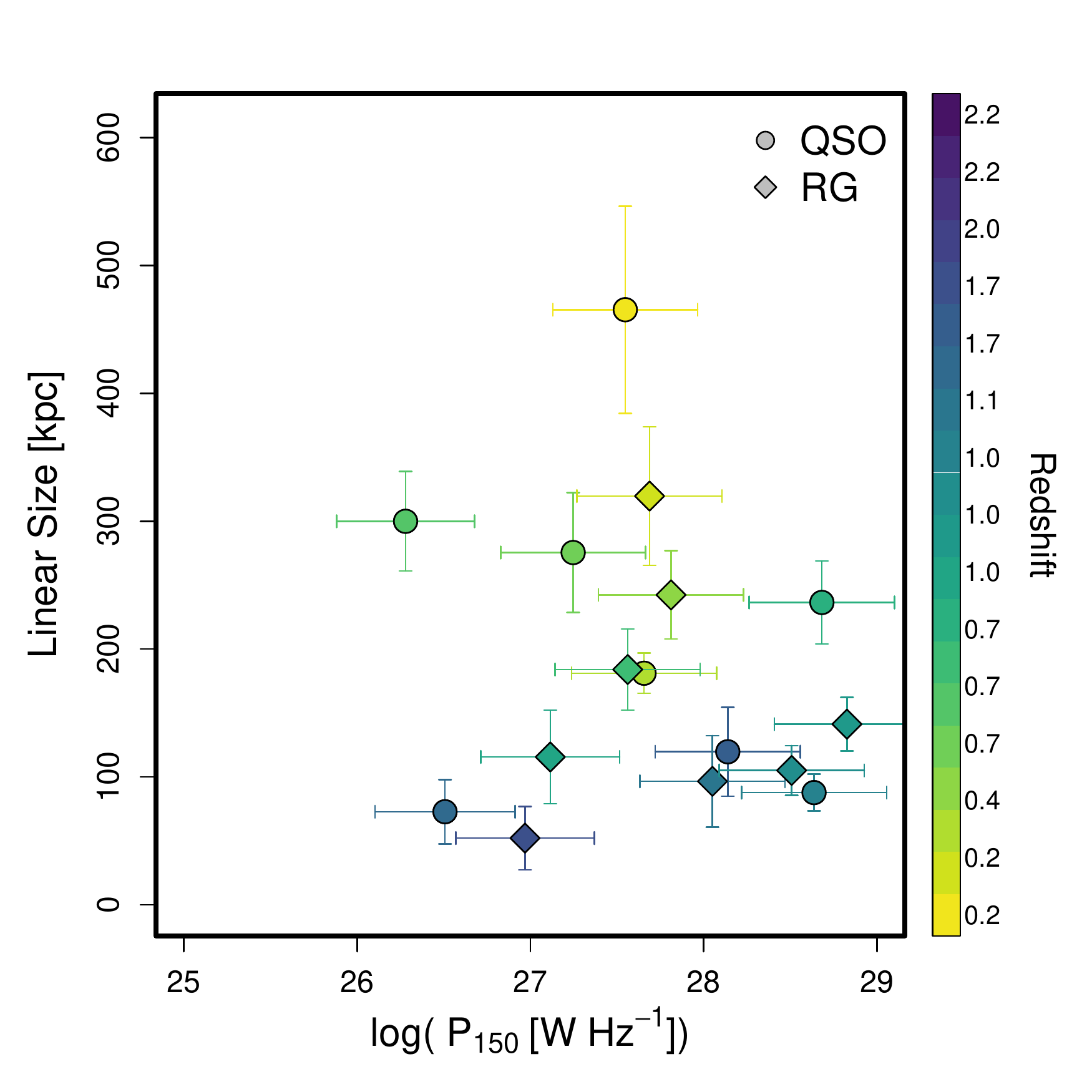}
\includegraphics[width=0.45\textwidth]{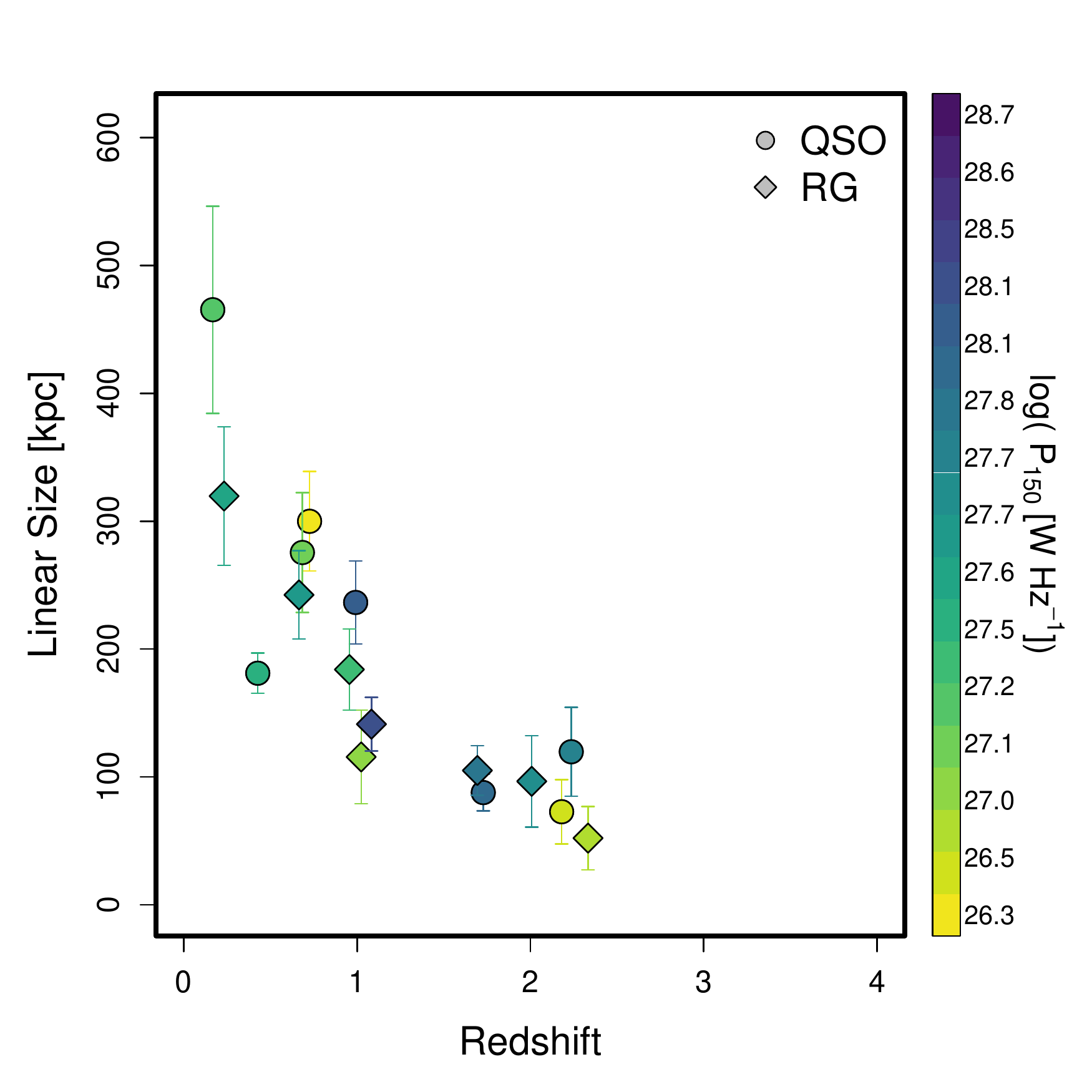}
\caption{\label{fig:f6}The linear sizes of quasars (circles) and radio galaxies (diamonds) plotted against power ($left$) and ($redshift$). The points represent the mean values from the binned data as listed in Table~\ref{tab:t1}. The data are The points are coloured according to the colour bar to the right of each plot and show the redshift ($left$) and power ($right$). }
\end{figure*}

We first examine the measured properties of radio galaxies and quasars by looking at how their projected linear sizes correlate with both power and redshift, by using the binned data. This is shown in Fig.~\ref{fig:f6}. There is no clear correlation between projected linear size and power. There does appear to be a correlation between linear size and redshift, with smaller objects found at higher redshifts. The anti-correlation of angular size with redshift is consistent with previous studies \citep{ker12,neeser95,wm74,miley68} which also find more compact radio sources at higher redshifts. 

To quantify the dependence of size on redshift we use the same Partial-rank analysis as described in \S 2.3 of \citet{neeser95}. Assuming a functional form of $D\propto(1+z)^{-n}$ where $D$ is linear size and $n$ is the `evolution strength,' we let $n$ vary in steps of 0.001 over the range $0\leq n \leq 3$. For each value of $n$ we multiplied the source sizes by $(1+z)^n$ and calculated the Partial-rank statistic for $D,z$ given $P$. The value of $n$ that produced a statistic of 0 was taken as the evolution strength, and the upper and lower errors correspond to where the statistic was $\pm1$. We find evolution strengths of $n=1.61^{+0.22}_{-0.22}$ for all radio galaxies and quasars considered together, $n=1.53^{+0.35}_{-0.35}$ for radio galaxies, and $n=1.64^{+0.22}_{-0.21}$ for quasars.  These values all agree within the uncertainties, indicating that quasars and radio galaxies have the same evolution strength. The values are also in excellent agreement with \citet{neeser95}, who find an evolution strength for galaxies and quasars of $n=1.71^{+0.40}_{-0.48}$ for a flat universe. 

We repeat the same exercise but looking at the Partial-rank statistic for $D,P$ given $z$. Using the functional form $D\propto P^m$, we multiply the sources sizes by $P^{-m}$ and vary $m$ in steps of $5\times10^{-5}$ from -1 to 1. The values of $m$ that produce statistics of 0 are consistent with $m=0$ for all samples. Specifically, for all sources we find $m=1\pm3\times10^{-5}$, for radio galaxies we find $m=1\pm1.5\times10^{-4}$, and for quasars we find $m=1\pm1.5\times10^{-5}$. We therefore conclude that there is no intrinsic relationship between power and size in this sample. 

Since an evolution of linear sizes with redshift exists, this could introduce a bias in the projected linear size \emph{ratios} if the mean redshifts of the radio galaxies and quasars are different. It is possible we could find a larger size ratio of radio galaxy to quasar linear sizes if the radio galaxies tend to be at lower redshifts than the quasars. We therefore use the evolution strengths calculated above to adjust the average linear sizes of each bin to the same redshift, in the following manner:

\begin{equation}
\label{eq:1}
D(z_1)=D(z_2) \frac{(1+z_1)^{-n} }{(1+z_2)^{-n}} .
\end{equation}

\begin{figure*}
\includegraphics[width=0.45\textwidth]{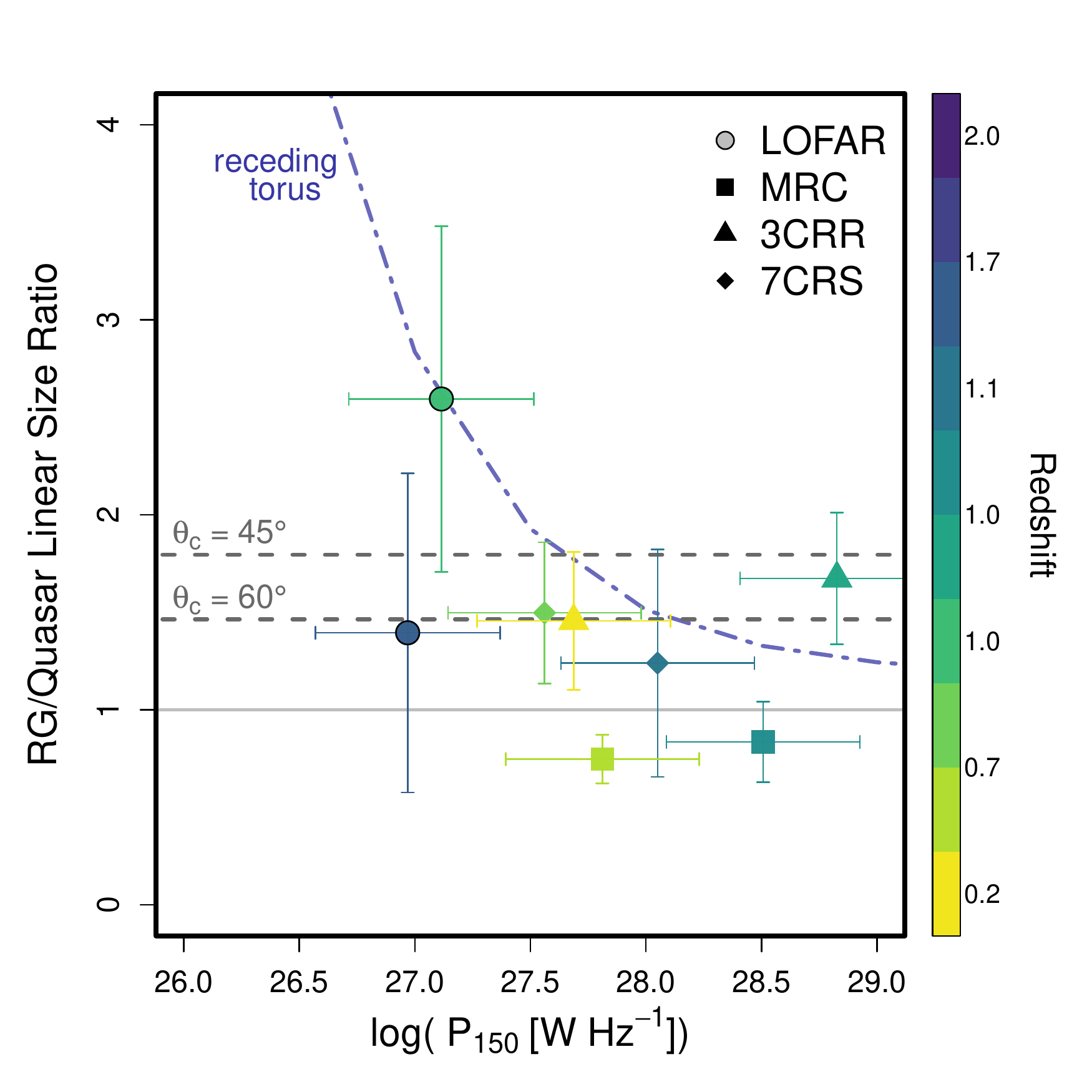}
\includegraphics[width=0.45\textwidth]{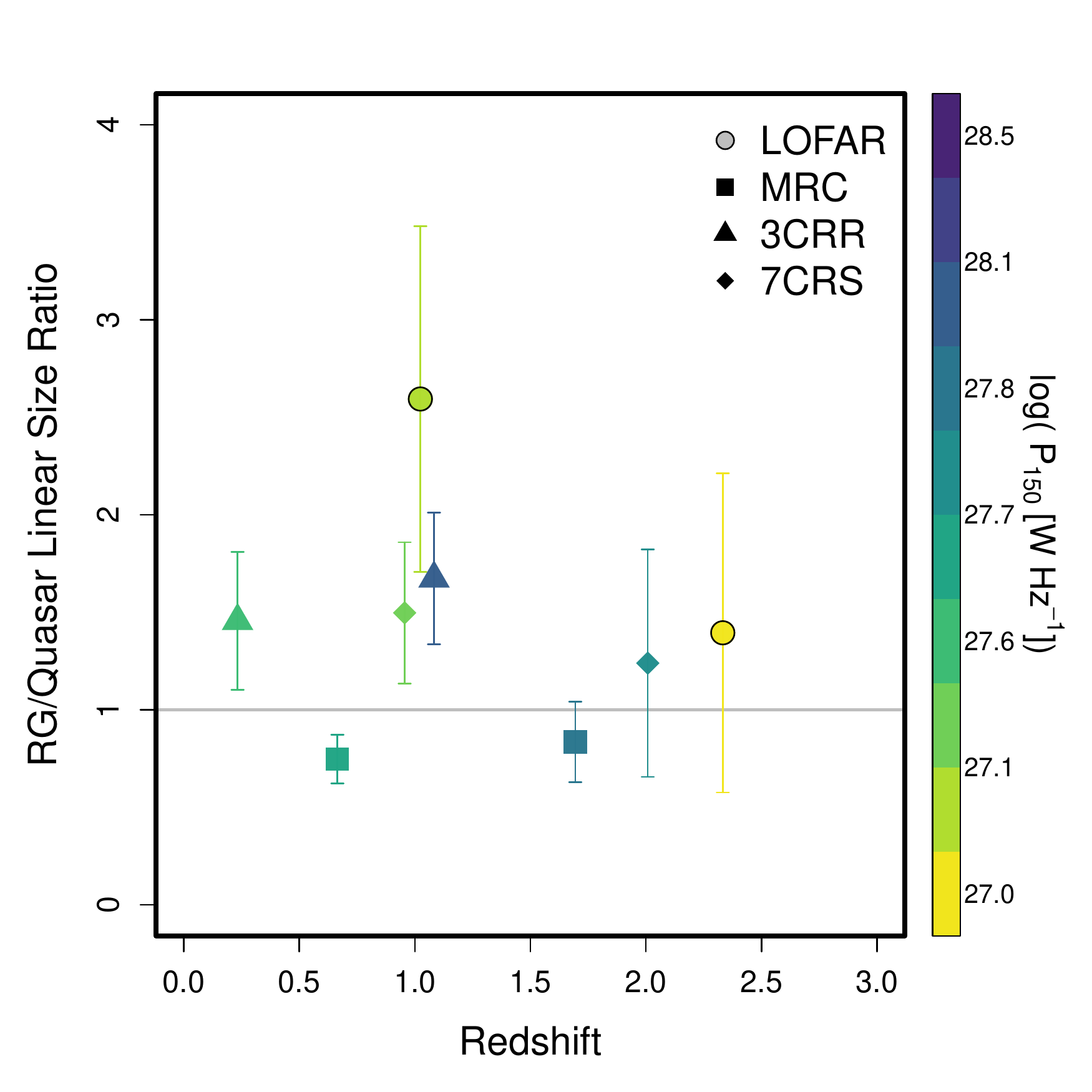}
\caption{\label{fig:f7}The radio galaxy to quasar linear size ratios plotted against power ($left$) and redshift ($right$). The symbol shapes represent the sample, with squares for the MRC, triangles for the 3CRR, diamonds for 7CRS, and circles for the LOFAR samples. The colour axes are redshift ($left$) and power ($right$). The horizontal gray lines indicate where the linear size ratio is unity. The linear size ratio vs. power plot also shows theoretical predictions for $(i)$ two constant torus opening angles, shown by the dark gray dashed lines, and $(ii)$ the receding torus model as described in \S~\ref{sec:s42}, shown by the blue dash-dotted lines.}
\end{figure*}

\begin{figure}
\includegraphics[width=0.45\textwidth]{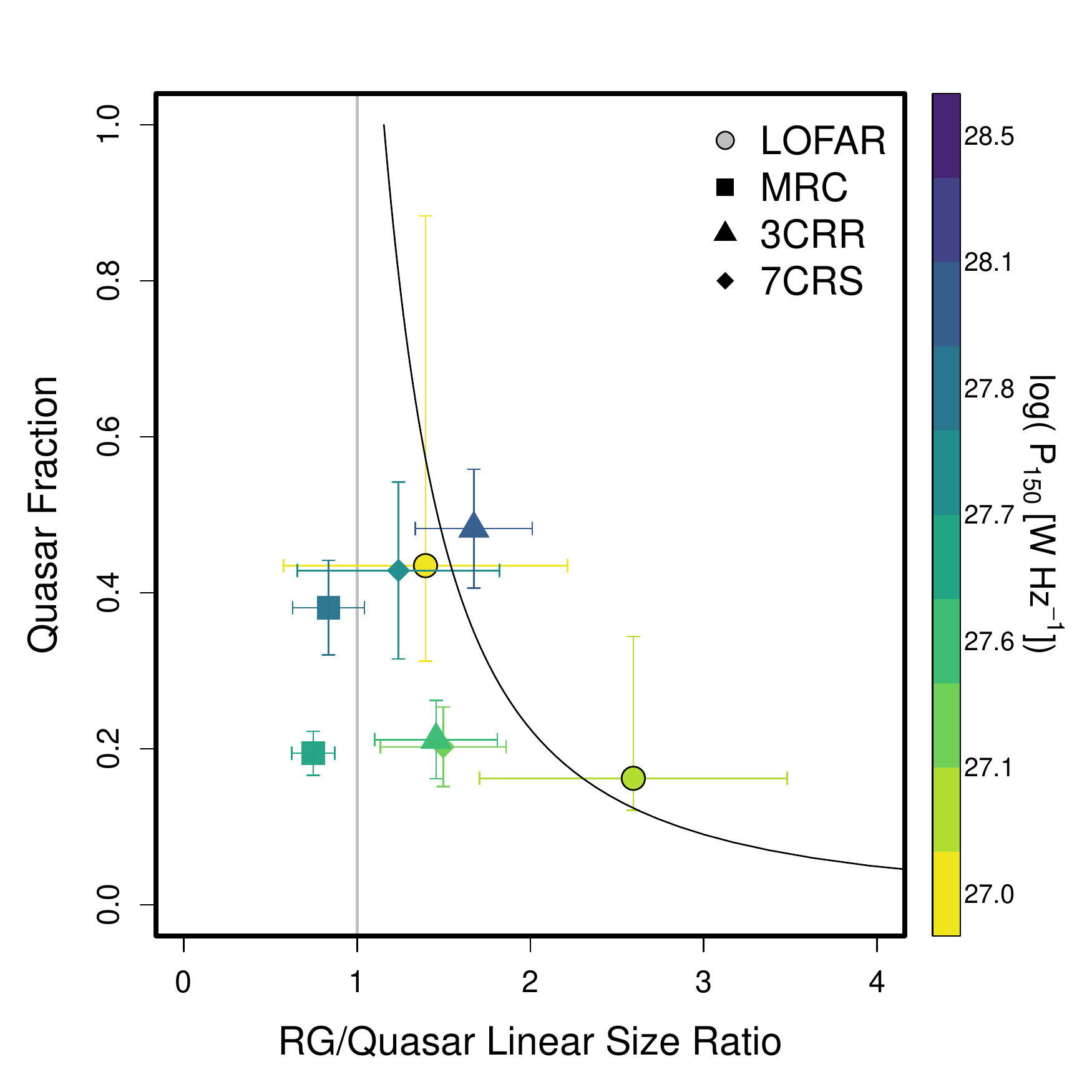}
\caption{\label{fig:f8}The quasar fraction vs. radio galaxy to quasar linear size ratio. Each sample is represented by symbols of different shapes as indicated in the legend. The horizontal gray line shows where the ratio between linear sizes of radio galaxies and quasars is unity. The curved black line is the prediction the relationship between observed quasar fraction and projected linear size ratio for a purely geometric orientation-based Unification model.  The color axis shows power. }
\end{figure}

This removes the possibility that the results will be biased if the average redshifts for the quasars and radio galaxies are different within a bin. We conservatively only do this for average values within a bin, as the propagation of errors across large redshift ranges will inflate the uncertainties in the linear sizes to be larger than the values themselves if the difference in redshift is too large. Additionally, the fit is not physically motivated and correcting across large redshift ranges is not advisable if, for example, flux limits contribute a significant amount to the redshift evolution of observed size. We correct both the radio galaxy and quasar projected linear sizes to the average redshift of \emph{all} sources in the bin, using the appropriate evolution strength for each sample. The values for average power, linear sizes, along with the number of sources, quasar fraction, and radio galaxy to quasar linear size ratio are listed in Tab.~\ref{tab:t1}. The linear size values are the corrected values, although the average redshift for the individual samples are listed for comparison. Figs.~\ref{fig:f7} and \ref{fig:f8} show linear size ratios calculated from the corrected linear sizes. 

We also use the redshift evolution to adjust the final results of \S~\ref{sec:s32}. This yields a corrected ratio of $2.0\pm0.3$. 

\begin{table*}
\centering
\caption{\label{tab:t1} Calculated values used for comparison of samples. Redshift ranges labelled `low' correspond to sources with redshifts less than or equal to the dividing redshift for each sample (LOFAR: $z=1.5$, 7CRS: $z=1.5$, 3CRR: $z=0.5$, MRC: $z=1$). The final column `Ref.' refers to the different samples:  L = LOFAR, 7 = 7CRS, 3 = 3CRR, M = MRC. Although the average redshifts, $\bar{z}$, are given for radio galaxies and quasars, the projected linear sizes (and therefore the ratios) are corrected to the average redshift for \emph{all} sources within a bin, using Eqn.~\ref{eq:1}.  }
\def\arraystretch{1.15}
\begin{tabular}{cccrrcccrrccll}
  \hline
  & \multicolumn{4}{c}{Radio Galaxies} & & \multicolumn{4}{c}{Quasars} & & \\ \cline{2-5} \cline{7-10} \\[-5pt]
  $z$ bin & $\bar{z}$ & Power & Linear size & No. &  &  $\bar{z}$ & Power& Linear size & No. & Quasar & Linear size & Ref. \\ 
  & &  [\wphz ] & [kpc]\phantom{LL}  & & & &  [\wphz ] & [kpc]\phantom{LL}  &  & Fraction & Ratio & \\
  \hline
  low & 0.726 & $1.9\pm0.01\times 10^{26}$ &  $300\pm39$ &   31 & & 1.02 & $1.3\pm0.01\times 10^{27}$ &  $116\pm37$ &    6 & $0.16^{+0.34}_{-0.12}$ & $2.59\pm0.89$ & L \\ 
   high & 2.18 & $3.2\pm0.05\times 10^{26}$ & $73\pm25$ &   13 &  & 2.33 & $9.3\pm0.07\times 10^{26}$ & $52\pm25$ &   10 & $0.43^{+0.88}_{-0.31}$ & $1.39\pm0.82$ & L \\ 
     low & 0.684 & $1.8\pm0.18\times 10^{27}$ &  $276\pm47$ &   63 &  & 0.956 & $3.6\pm0.36\times 10^{27}$ &  $184\pm32$ &   16 & $0.20^{+0.41}_{-0.12}$ & $1.50\pm0.36$ & 7 \\ 
   high & 2.24 & $1.4\pm0.14\times 10^{28}$ &  $120\pm35$ &   20 &  & 2.01 & $1.1\pm0.11\times 10^{28}$ & $97\pm36$ &   15 & $0.43^{+0.85}_{-0.28}$ & $1.24\pm0.58$ & 7 \\ 
     low & 0.167 & $3.5\pm0.35\times 10^{27}$ &  $465\pm81$ &   67 &  & 0.233 & $4.9\pm0.49\times 10^{27}$ &  $320\pm54$ &   18 & $0.21^{+0.43}_{-0.12}$ & $1.46\pm0.35$ & 3 \\ 
   high & 0.992 & $4.8\pm0.48\times 10^{28}$ &  $236\pm33$ &   44  & & 1.08 & $6.7\pm0.67\times 10^{28}$ &  $141\pm21$ &   41 & $0.48^{+0.93}_{-0.25}$ & $1.67\pm0.34$ & 3 \\ 
     low & 0.427 & $4.5\pm0.45\times 10^{27}$ &  $181\pm16$ &  199  & & 0.665 & $6.5\pm0.65\times 10^{27}$ &  $242\pm35$ &   48 & $0.19^{+0.38}_{-0.09}$ & $0.75\pm0.12$ & M \\ 
     high & 1.73 & $4.3\pm0.43\times 10^{28}$ & $88\pm14$ &   65  & & 1.69 & $3.2\pm0.32\times 10^{28}$ &  $105\pm19$ &   40 & $0.38^{+0.74}_{-0.19}$ & $0.84\pm0.21$ & M \\ 
   \hline
\end{tabular}
\end{table*}

We then investigate how the ratio between radio galaxy and quasar projected linear sizes evolves with power and with redshift, see Fig.~\ref{fig:f7}. The LOFAR, 3CRR, and 7CRS samples generally show size ratios larger than unity, while the MRC sample shows size ratios below unity. The errors are large enough that the high-redshift bins for the LOFAR and 7CRS samples could be consistent with unity. There does not appear to be a trend with either power or redshift. 
This indicates that the linear size ratio remains the same for a large range of powers and out to high redshifts. 
The fact that the linear size ratio does not clearly evolve with either power or redshift implies that the populations of radio galaxies and quasars \emph{always} have the same relative sizes, regardless of power or redshift evolution. 

\citet{lawrence91} has proposed that observed increasing quasar fractions with power could arise naturally from a `receding torus' model. In this model, the inner radius of the dusty torus (and hence the opening angle, $\theta_c$) increases as the temperatures achieved via radiation from increasingly powerful AGN can sublimate dust at larger distances. Using the formulation in \S~4.3 of \citet{willott00}, we can directly relate the predicted linear size ratio to the radio power using the authors' functional form in Eqn. (1):
\begin{equation}
f_q =  1 - \left(	1 + \frac{P_{150}}{P^*_{150}}\textrm{tan}^2\theta_0 \right)^{-0.5} , 
\end{equation}
replacing $L$ with $P_{150}$. We calculate the normalisations, $P^*_{150}$ and $\theta_0$, from the averages of the power and size ratios given in Table~\ref{tab:t2}. We use the size ratios rather than the quasar fractions to reduce the potential for bias in the quasar fraction (see \S~\ref{sec:s24}). The normalisation values are $P^*_{150} = 1.58\times10^{28}\,$\wphz\ and $\theta_0=62.2^{\circ}$. The quasar fractions then directly define the predicted linear size ratios. These are plotted in Fig.~\ref{fig:f7} (left panel). We also plot the linear size ratios predicted from  assuming that the receding torus model does not hold, for two constant opening angles ($\theta_c=45^{\circ}$ and $\theta_c=60^{\circ}$). Both constant opening angles are consistent within the uncertainties of all data points from the LOFAR, 3CRR, and 7CRS samples. The receding torus model describes most of these data points reasonably well, and cannot be ruled out. The largest difference between the two models is for lower powers, and future work expanding the LOFAR sample to reduce the uncertainties will be important. 

Finally we investigate how the quasar fraction depends on the radio galaxy to quasar linear size ratio. The results are shown in Fig.~\ref{fig:f8}. For a purely geometric orientation-based unification of radio galaxies and quasars, the observed quasar fraction will define the projected linear size. Apart from the MRC sample and the two low-redshift bins of the 3CRR and 7CRS samples, the data in general are consistent with the theoretical predictions. The predictions assume no intrinsic size distribution, which can artificially inflate the linear size ratio by about 10 per cent \citep[estimated from Fig. 3 of][]{dipompeo13}. This effect is small compared to the uncertainties we calculate, it is not likely to change our results. The high-redshift LOFAR bin could have a biased quasar fraction as discussed in \S~\ref{sec:s24}, but even if the quasar fraction in the high-redshift bin were reduced by a factor of two (bringing it in agreement with the low-redshift bin), the data point in Fig.~\ref{fig:f8} would still be consistent with the theoretical prediction, within the uncertainties.
 
\section{Discussion}
\label{sec:s5}
For the discussion, we first consider how the radio data can be interpreted given an orientation-based unification scenario. We then discuss the possibility that evolution rather than orientation is the dominant effect that explains the data. 

\subsection{Orientation Interpretation}
\label{sec:s51}
First we consider a scheme where the observed fraction of quasars depends \emph{only} on viewing angle, which is supported by other observational evidence. For example, \citet{antonucci85} used high-resolution observations of blazars to show that they are consistent with being normal radio galaxies viewed along the jet axis. Observed differences in depolarisation of radio lobes (the Laing-Garrington effect) also indicate orientation effects, as the approaching (receding) lobe will appear less (more) depolarised due to differential Faraday rotation in the ambient medium along the line of sight \citep{gl88}. In some cases, this can cause sources close to the line of sight (i.e., quasars) to appear one-sided as the receding jet drops below the sensitivity limit. If we assume that all quasars in our LOFAR sample are one-sided and multiply their sizes by a factor of two to account for this, we still find a linear size ratio of radio galaxies to quasars of $1.53\pm0.48$. This is still above unity and therefore consistent with an orientation scheme. This is the \emph{smallest} that the ratio could possibly be, as some quasars have clear double structure. 

The unification of HERGs via orientation predicts several key observable characteristics with which we can compare our results. First, the projected linear sizes of quasars should on average be smaller than the projected linear sizes of radio galaxies. The new LOFAR data are consistent with this, as we found a linear size ratio of 2.0$\pm$0.3 (for radio galaxies to quasars, after correcting for redshift evolution). For the LOFAR sample we found a division angle between radio galaxies and quasars of $39.2^{+9.2}_{-6.1}$ degrees, which is in agreement with the value of 44.4$^{\circ}$ found for the original 3CRR sample \citep{barthel89,willott00}. That would mean that quasars have their radio jets oriented between 0$^{\circ}$ (line of sight) and $39^{\circ}$, while radio galaxies have jets oriented between $39^{\circ}$ and 90$^{\circ}$ (plane of sky). This is similar to values found from other studies, which use other methods to estimate the critical division angle between quasars and radio galaxies, based on parameters like core dominance or emission line properties. In general, quasars (sometimes called `broad line objects') have higher values of core dominance, which have been used to estimate division angles of 50$\pm5$ degrees \citep{baldi13}, 60$\pm10$ degrees \citep{ma16}, or constrained to between 10 and 80 degrees \citep{aars05}. 

Second, beamed radio sources with flat spectra should be oriented close to the line of sight, and therefore should also be smaller on average than steep-spectrum sources.  The LOFAR data is also consistent with this scenario, and the linear size ratio is 4.4$\pm$1.4, even larger than the linear size ratio for radio galaxies/quasars. The angle of division between flat and steep spectrum sources is $35.3^{+5.2}_{-5.9}$ degrees. This is consistent with the idea that the flux densities of  flat-spectrum sources are dominated by beamed radio jets that are oriented close to the line of sight. The fact that this angle is smaller than the division between radio galaxies and quasars is consistent with an orientation-only scheme where there are fewer sources that are beamed than are identified as quasars, if the opening angle of the torus is what determines whether or not a quasar is observed.

Finally, the quasar fraction should be directly correlated with the linear size ratio if \emph{only} orientation is responsible for whether or not we observe a quasar. We find that the LOFAR data is entirely consistent with the predictions (see Fig.~\ref{fig:f8}). The high-redshift bins of both the 7CRS and 3CRR samples are also consistent with this prediction, while the low-redshift bins lie slightly below the prediction. The sizes of the samples limited the analysis to only two redshift bins per sample. With LoTSS we will be able to fill enough of the $P-z$ plane to refine the redshift bins and investigate this inconsistency between the low- and high-redshift bins. The MRC sample is not at all consistent with the orientation-only predictions. This is the only sample of the four we investigated that shows no difference between the sizes of radio galaxies and quasars. Since we treat the samples separately, any systematic offset in size measurements should affect both radio galaxies and quasars in the MRC sample in the same manner, and the \emph{ratio} of the two would still be robust. 
If there is a population of radio sources that do not participate in the orientation-based unification scheme, the higher selection frequency of the MRC may be impacted to a larger extent than the other surveys. We can find no evidence of this in the available information, but it is possible that the identification and removal of LERGs from the MRC sample could be instructive. 

Overall, we find that the LOFAR results are consistent with an orientation-based unification scenario, as are the high-redshift data from the 3CRR and 7CRS samples. 

\subsection{Evolutionary Interpretation}
\label{sec:s52}
Another possibility is that radio galaxies and quasars are linked through an evolutionary scheme rather than by orientation alone. 
In such a scheme, radio jets would be triggered when quasars become active. The radio jets would grow and finally when the quasar reaches an inactive state the source would be classified as a radio galaxy (still with AGN signatures, but lack of emission from an accretion disk). The measured quasar fraction in this case would be interpreted as the fraction of time a source spends as an active quasar, and the linear size ratio would depend on the expansion rate of the radio source. Radio galaxies would be larger than quasars because they are older and have had more time to expand, but this does not account for recurrent quasar activity, as the projected linear size would be correlated with the \emph{first} quasar activity. In the LOFAR sample the sizes of radio galaxies are on average larger than the sizes of quasars, which is consistent with this evolutionary scheme. However, evolution alone cannot explain observational effects like the Laing-Garrington effect or the presence of scattered quasar light in radio galaxies \citep[e.g.,][]{jts98}. 

If an evolutionary scheme holds, the fact that higher quasar fractions are seen for the higher redshift bins would mean that either higher redshift sources spend a longer portion of their lives as quasars before becoming radio galaxies, or that there are simply more young quasars than old radio galaxies at high redshift. If it is true that higher redshift sources spend longer portions of their lives as quasars, the radio jets would have to grow more slowly to be consistent with Fig.~\ref{fig:f7}, where we see no change in the linear size ratio with redshift. Slower growth of radio jets could be due to the higher density ambient medium expected at high redshift, but it would not change the amount of time that a quasar is active. If there are simply more young quasars at higher redshift, we would expect to see an evolution with redshift of the linear size ratios between quasars and radio galaxies, which we do not observe. This could possibly be rectified if \emph{all} sources (quasars and radio galaxies) at high redshift were younger than those at low redshift. Of course evolutionary effects will play a role -- but apparently not a dominant one.

\section{Conclusions}
\label{sec:s6}
In this paper we used a new LOFAR survey of the Bo\"{o}tes field to show that the projected linear sizes of steep-spectrum radio sources are on average $4.4\pm 1.4$ times larger than those of flat-spectrum radio sources. This is consistent with an orientation scheme for radio jets, where beamed flat-spectrum radio sources lie closer to the line of sight and therefore have smaller projected sizes. 

We have also shown that for radio galaxies and quasars in the LOFAR survey, as identified by AGES criteria, the projected linear sizes of radio galaxies are on average $3.1\pm 1.0$ times larger than those of quasars, which becomes $2.0\pm0.3$ after correcting for the redshift evolution of linear sizes. This is also consistent with an orientation-based unification scheme, where the presence of a dusty obscuring torus prevents the identification of a quasar unless the radio jets are preferentially aligned closer to the line of sight. 

When combining the new LOFAR measurements with previous surveys and separating each sample into low and high redshift bins, we find no clear trend between the linear size ratio of radio galaxies to quasars and redshift or power. The lack of a clear trend between the linear size ratio and redshift suggests that the populations of radio galaxies and quasars \emph{always} have the same sizes relative to each other. The lack of a trend with redshift is consistent with orientation, or else an evolutionary scheme where the sizes of radio galaxies and quasars are evolving in exactly the same manner at exactly the same rate. 

When comparing theoretical predictions of orientation-based Unification models for the relation between quasar fraction and linear size ratio to the observations, the LOFAR data are consistent with the predictions. Other low-frequency surveys are in agreement for their high-redshift bins, while the MRC sample, which is measured at a higher frequency, consistently shows that quasars are on average larger than radio galaxies. 
The comparison between the linear size ratios and power shows that the data (aside from the MRC sample) are consistent with a constant opening angle for sources of all powers, but the `receding torus' model cannot be ruled out at the present time. The largest difference between the models is for lower-power sources, and expanding the LOFAR sample to reduce the uncertainties will be crucial to distinguishing between these two models. 

Ultimately the LOFAR Tier 1 survey will cover the Northern sky above declination 0 degrees, providing millions of radio sources. Spectroscopic redshifts and host galaxy identifications will be provided by a survey with the William Herschel Telescope Enhanced Area Velocity Explorer (WEAVE) which goes online March 2018. The WEAVE-LOFAR survey \citep{weave} is dedicated to providing this information for $\sim10^6$ LOFAR-detected sources in the Tier 1 survey. With this information, and updated theoretical models that allow us to account for intrinsic size distributions (Saxena et al, submitted) we will be able to break the degeneracy between orientation and evolutionary effects in unification schemes.

\section*{Acknowledgements}
We thank the anonymous referee for their useful comments on this manuscript. LKM acknowledges financial support from NWO Top LOFAR project, project n. 614.001.006. WLW acknowledges support from the UK Science and Technology Facilities Council [ST/M001008/1]. PNB is grateful for support from the UK STFC via grant ST/M001229/1. MJJ acknowledges support from UK Science and Technology Facilities Council and the South African SKA Project. EKM acknowledges support from the Australian Research Council Centre of Excellence for All-sky Astrophysics (CAASTRO), through project number CE110001020. HR and KJD acknowledge support from the ERC Advanced Investigator programme NewClusters 321271.

\label{lastpage}


\begin{thebibliography}{}
\makeatletter
\relax
\def\mn@urlcharsother{\let\do\@makeother \do\$\do\&\do\#\do\^\do\_\do\%\do\~}
\def\mn@doi{\begingroup\mn@urlcharsother \@ifnextchar [ {\mn@doi@}
  {\mn@doi@[]}}
\def\mn@doi@[#1]#2{\def\@tempa{#1}\ifx\@tempa\@empty \href
  {http://dx.doi.org/#2} {doi:#2}\else \href {http://dx.doi.org/#2} {#1}\fi
  \endgroup}
\def\mn@eprint#1#2{\mn@eprint@#1:#2::\@nil}
\def\mn@eprint@arXiv#1{\href {http://arxiv.org/abs/#1} {{\tt arXiv:#1}}}
\def\mn@eprint@dblp#1{\href {http://dblp.uni-trier.de/rec/bibtex/#1.xml}
  {dblp:#1}}
\def\mn@eprint@#1:#2:#3:#4\@nil{\def\@tempa {#1}\def\@tempb {#2}\def\@tempc
  {#3}\ifx \@tempc \@empty \let \@tempc \@tempb \let \@tempb \@tempa \fi \ifx
  \@tempb \@empty \def\@tempb {arXiv}\fi \@ifundefined
  {mn@eprint@\@tempb}{\@tempb:\@tempc}{\expandafter \expandafter \csname
  mn@eprint@\@tempb\endcsname \expandafter{\@tempc}}}

\bibitem[\protect\citeauthoryear{{Aars}, {Hough}, {Yu}, {Linick}, {Beyer},
  {Vermeulen}  \& {Readhead}}{{Aars} et~al.}{2005}]{aars05}
{Aars} C.~E.,  {Hough} D.~H.,  {Yu} L.~H.,  {Linick} J.~P.,  {Beyer} P.~J.,
  {Vermeulen} R.~C.,   {Readhead} A.~C.~S.,  2005, \mn@doi [\aj]
  {10.1086/430530}, \href {http://adsabs.harvard.edu/abs/2005AJ....130...23A}
  {130, 23}

\bibitem[\protect\citeauthoryear{{Antonucci}}{{Antonucci}}{1982}]{antonucci82}
{Antonucci} R.~R.~J.,  1982, \mn@doi [\nat] {10.1038/299605a0}, \href
  {http://adsabs.harvard.edu/abs/1982Natur.299..605A} {299, 605}

\bibitem[\protect\citeauthoryear{{Antonucci}}{{Antonucci}}{1984}]{antonucci84}
{Antonucci} R.~R.~J.,  1984, \mn@doi [\apj] {10.1086/161816}, \href
  {http://adsabs.harvard.edu/abs/1984ApJ...278..499A} {278, 499}

\bibitem[\protect\citeauthoryear{{Antonucci}}{{Antonucci}}{1993}]{antonucci93}
{Antonucci} R.,  1993, \mn@doi [\araa] {10.1146/annurev.aa.31.090193.002353},
  \href {http://adsabs.harvard.edu/abs/1993ARA%26A..31..473A} {31, 473}

\bibitem[\protect\citeauthoryear{{Antonucci} \& {Ulvestad}}{{Antonucci} \&
  {Ulvestad}}{1985}]{antonucci85}
{Antonucci} R.~R.~J.,  {Ulvestad} J.~S.,  1985, \mn@doi [\apj]
  {10.1086/163284}, \href {http://adsabs.harvard.edu/abs/1985ApJ...294..158A}
  {294, 158}

\bibitem[\protect\citeauthoryear{{Baker}, {Hunstead}, {Kapahi}  \&
  {Subrahmanya}}{{Baker} et~al.}{1999}]{MRCpart4}
{Baker} J.~C.,  {Hunstead} R.~W.,  {Kapahi} V.~K.,   {Subrahmanya} C.~R.,
  1999, \mn@doi [\apjs] {10.1086/313209}, \href
  {http://adsabs.harvard.edu/abs/1999ApJS..122...29B} {122, 29}

\bibitem[\protect\citeauthoryear{{Baldi}, {Capetti}, {Buttiglione}, {Chiaberge}
   \& {Celotti}}{{Baldi} et~al.}{2013}]{baldi13}
{Baldi} R.~D.,  {Capetti} A.,  {Buttiglione} S.,  {Chiaberge} M.,   {Celotti}
  A.,  2013, \mn@doi [\aap] {10.1051/0004-6361/201322842}, \href
  {http://adsabs.harvard.edu/abs/2013A%26A...560A..81B} {560, A81}

\bibitem[\protect\citeauthoryear{{Barthel}}{{Barthel}}{1989}]{barthel89}
{Barthel} P.~D.,  1989, \mn@doi [\apj] {10.1086/167038}, \href
  {http://adsabs.harvard.edu/abs/1989ApJ...336..606B} {336, 606}

\bibitem[\protect\citeauthoryear{{Barthel}, {Hooimeyer}, {Schilizzi}, {Miley}
  \& {Preuss}}{{Barthel} et~al.}{1989}]{barthel89a}
{Barthel} P.~D.,  {Hooimeyer} J.~R.,  {Schilizzi} R.~T.,  {Miley} G.~K.,
  {Preuss} E.,  1989, \mn@doi [\apj] {10.1086/167037}, \href
  {http://adsabs.harvard.edu/abs/1989ApJ...336..601B} {336, 601}

\bibitem[\protect\citeauthoryear{{Best} \& {Heckman}}{{Best} \&
  {Heckman}}{2012}]{bh12}
{Best} P.~N.,  {Heckman} T.~M.,  2012, \mn@doi [\mnras]
  {10.1111/j.1365-2966.2012.20414.x}, \href
  {http://adsabs.harvard.edu/abs/2012MNRAS.421.1569B} {421, 1569}

\bibitem[\protect\citeauthoryear{{Best}, {R{\"o}ttgering}  \& {Lehnert}}{{Best}
  et~al.}{1999}]{blr99}
{Best} P.~N.,  {R{\"o}ttgering} H.~J.~A.,   {Lehnert} M.~D.,  1999, \mn@doi
  [\mnras] {10.1046/j.1365-8711.1999.02984.x}, \href
  {http://adsabs.harvard.edu/abs/1999MNRAS.310..223B} {310, 223}

\bibitem[\protect\citeauthoryear{{Brammer}, {van Dokkum}  \& {Coppi}}{{Brammer}
  et~al.}{2008}]{brammer08}
{Brammer} G.~B.,  {van Dokkum} P.~G.,   {Coppi} P.,  2008, \mn@doi [\apj]
  {10.1086/591786}, \href {http://adsabs.harvard.edu/abs/2008ApJ...686.1503B}
  {686, 1503}

\bibitem[\protect\citeauthoryear{{Brown}, {Dey}, {Jannuzi}, {Brand}, {Benson},
  {Brodwin}, {Croton}  \& {Eisenhardt}}{{Brown} et~al.}{2007}]{brown07}
{Brown} M.~J.~I.,  {Dey} A.,  {Jannuzi} B.~T.,  {Brand} K.,  {Benson} A.~J.,
  {Brodwin} M.,  {Croton} D.~J.,   {Eisenhardt} P.~R.,  2007, \mn@doi [\apj]
  {10.1086/509652}, \href {http://adsabs.harvard.edu/abs/2007ApJ...654..858B}
  {654, 858}

\bibitem[\protect\citeauthoryear{{Brown} et~al.,}{{Brown}
  et~al.}{2008}]{brown08}
{Brown} M.~J.~I.,  et~al., 2008, \mn@doi [\apj] {10.1086/589538}, \href
  {http://adsabs.harvard.edu/abs/2008ApJ...682..937B} {682, 937}

\bibitem[\protect\citeauthoryear{{Brown} et~al.,}{{Brown}
  et~al.}{2014}]{brown14}
{Brown} M.~J.~I.,  et~al., 2014, \mn@doi [\apjs] {10.1088/0067-0049/212/2/18},
  \href {http://adsabs.harvard.edu/abs/2014ApJS..212...18B} {212, 18}

\bibitem[\protect\citeauthoryear{{Cohen}, {Ogle}, {Tran}, {Goodrich}  \&
  {Miller}}{{Cohen} et~al.}{1999}]{cohen99}
{Cohen} M.~H.,  {Ogle} P.~M.,  {Tran} H.~D.,  {Goodrich} R.~W.,   {Miller}
  J.~S.,  1999, \mn@doi [\aj] {10.1086/301074}, \href
  {http://adsabs.harvard.edu/abs/1999AJ....118.1963C} {118, 1963}

\bibitem[\protect\citeauthoryear{{DiPompeo}, {Runnoe}, {Myers}  \&
  {Boroson}}{{DiPompeo} et~al.}{2013}]{dipompeo13}
{DiPompeo} M.~A.,  {Runnoe} J.~C.,  {Myers} A.~D.,   {Boroson} T.~A.,  2013,
  \mn@doi [\apj] {10.1088/0004-637X/774/1/24}, \href
  {http://adsabs.harvard.edu/abs/2013ApJ...774...24D} {774, 24}

\bibitem[\protect\citeauthoryear{{Evans}, {Worrall}, {Hardcastle}, {Kraft}  \&
  {Birkinshaw}}{{Evans} et~al.}{2006}]{evans06}
{Evans} D.~A.,  {Worrall} D.~M.,  {Hardcastle} M.~J.,  {Kraft} R.~P.,
  {Birkinshaw} M.,  2006, \mn@doi [\apj] {10.1086/500658}, \href
  {http://adsabs.harvard.edu/abs/2006ApJ...642...96E} {642, 96}

\bibitem[\protect\citeauthoryear{{Fanaroff} \& {Riley}}{{Fanaroff} \&
  {Riley}}{1974}]{fr74}
{Fanaroff} B.~L.,  {Riley} J.~M.,  1974, \mn@doi [\mnras]
  {10.1093/mnras/167.1.31P}, \href
  {http://adsabs.harvard.edu/abs/1974MNRAS.167P..31F} {167, 31P}

\bibitem[\protect\citeauthoryear{{Garrington}, {Leahy}, {Conway}  \&
  {Laing}}{{Garrington} et~al.}{1988}]{gl88}
{Garrington} S.~T.,  {Leahy} J.~P.,  {Conway} R.~G.,   {Laing} R.~A.,  1988,
  \mn@doi [\nat] {10.1038/331147a0}, \href
  {http://esoads.eso.org/abs/1988Natur.331..147G} {331, 147}

\bibitem[\protect\citeauthoryear{{Gehrels}}{{Gehrels}}{1986}]{gehrels86}
{Gehrels} N.,  1986, \mn@doi [\apj] {10.1086/164079}, \href
  {http://adsabs.harvard.edu/abs/1986ApJ...303..336G} {303, 336}

\bibitem[\protect\citeauthoryear{{Grimes}, {Rawlings}  \& {Willott}}{{Grimes}
  et~al.}{2004}]{grimes04}
{Grimes} J.~A.,  {Rawlings} S.,   {Willott} C.~J.,  2004, \mn@doi [\mnras]
  {10.1111/j.1365-2966.2004.07510.x}, \href
  {http://adsabs.harvard.edu/abs/2004MNRAS.349..503G} {349, 503}

\bibitem[\protect\citeauthoryear{{Hardcastle}, {Evans}  \&
  {Croston}}{{Hardcastle} et~al.}{2006}]{hardcastle06}
{Hardcastle} M.~J.,  {Evans} D.~A.,   {Croston} J.~H.,  2006, \mn@doi [\mnras]
  {10.1111/j.1365-2966.2006.10615.x}, \href
  {http://adsabs.harvard.edu/abs/2006MNRAS.370.1893H} {370, 1893}

\bibitem[\protect\citeauthoryear{{Hardcastle}, {Evans}  \&
  {Croston}}{{Hardcastle} et~al.}{2007}]{hardcastle07}
{Hardcastle} M.~J.,  {Evans} D.~A.,   {Croston} J.~H.,  2007, \mn@doi [\mnras]
  {10.1111/j.1365-2966.2007.11572.x}, \href
  {http://adsabs.harvard.edu/abs/2007MNRAS.376.1849H} {376, 1849}

\bibitem[\protect\citeauthoryear{{Ho}}{{Ho}}{2008}]{ho08}
{Ho} L.~C.,  2008, \mn@doi [\araa] {10.1146/annurev.astro.45.051806.110546},
  \href {http://adsabs.harvard.edu/abs/2008ARA%26A..46..475H} {46, 475}

\bibitem[\protect\citeauthoryear{{Jackson}, {Tadhunter}  \& {Sparks}}{{Jackson}
  et~al.}{1998}]{jts98}
{Jackson} N.,  {Tadhunter} C.,   {Sparks} W.~B.,  1998, \mn@doi [\mnras]
  {10.1046/j.1365-8711.1998.02008.x}, \href
  {http://adsabs.harvard.edu/abs/1998MNRAS.301..131J} {301, 131}

\bibitem[\protect\citeauthoryear{{Jannuzi} \& {Dey}}{{Jannuzi} \&
  {Dey}}{1999}]{jd99}
{Jannuzi} B.~T.,  {Dey} A.,  1999, in {Bunker} A.~J.,  {van Breugel} W.~J.~M.,
  eds,  Astronomical Society of the Pacific Conference Series Vol. 193, The
  Hy-Redshift Universe: Galaxy Formation and Evolution at High Redshift. p.~258

\bibitem[\protect\citeauthoryear{Janssen}{Janssen}{2017}]{janssen17}
Janssen R.,  2017, PhD thesis, Leiden University

\bibitem[\protect\citeauthoryear{{Kapahi}, {Athreya}, {van Breugel}, {McCarthy}
   \& {Subrahmanya}}{{Kapahi} et~al.}{1998a}]{MRCpart2}
{Kapahi} V.~K.,  {Athreya} R.~M.,  {van Breugel} W.,  {McCarthy} P.~J.,
  {Subrahmanya} C.~R.,  1998a, \mn@doi [\apjs] {10.1086/313144}, \href
  {http://adsabs.harvard.edu/abs/1998ApJS..118..275K} {118, 275}

\bibitem[\protect\citeauthoryear{{Kapahi}, {Athreya}, {Subrahmanya}, {Baker},
  {Hunstead}, {McCarthy}  \& {van Breugel}}{{Kapahi} et~al.}{1998b}]{MRCpart3}
{Kapahi} V.~K.,  {Athreya} R.~M.,  {Subrahmanya} C.~R.,  {Baker} J.~C.,
  {Hunstead} R.~W.,  {McCarthy} P.~J.,   {van Breugel} W.,  1998b, \mn@doi
  [\apjs] {10.1086/313145}, \href
  {http://adsabs.harvard.edu/abs/1998ApJS..118..327K} {118, 327}

\bibitem[\protect\citeauthoryear{{Ker}, {Best}, {Rigby}, {R{\"o}ttgering}  \&
  {Gendre}}{{Ker} et~al.}{2012}]{ker12}
{Ker} L.~M.,  {Best} P.~N.,  {Rigby} E.~E.,  {R{\"o}ttgering} H.~J.~A.,
  {Gendre} M.~A.,  2012, \mn@doi [\mnras] {10.1111/j.1365-2966.2011.20235.x},
  \href {http://adsabs.harvard.edu/abs/2012MNRAS.420.2644K} {420, 2644}

\bibitem[\protect\citeauthoryear{{Kochanek} et~al.,}{{Kochanek}
  et~al.}{2012}]{kochanek12}
{Kochanek} C.~S.,  et~al., 2012, \mn@doi [\apjs] {10.1088/0067-0049/200/1/8},
  \href {http://adsabs.harvard.edu/abs/2012ApJS..200....8K} {200, 8}

\bibitem[\protect\citeauthoryear{{Lacy}, {Rawlings}, {Hill}, {Bunker},
  {Ridgway}  \& {Stern}}{{Lacy} et~al.}{1999}]{lacy99}
{Lacy} M.,  {Rawlings} S.,  {Hill} G.~J.,  {Bunker} A.~J.,  {Ridgway} S.~E.,
  {Stern} D.,  1999, \mn@doi [\mnras] {10.1046/j.1365-8711.1999.02790.x}, \href
  {http://adsabs.harvard.edu/abs/1999MNRAS.308.1096L} {308, 1096}

\bibitem[\protect\citeauthoryear{{Laing}, {Riley}  \& {Longair}}{{Laing}
  et~al.}{1983}]{3CRR}
{Laing} R.~A.,  {Riley} J.~M.,   {Longair} M.~S.,  1983, \mn@doi [\mnras]
  {10.1093/mnras/204.1.151}, \href
  {http://adsabs.harvard.edu/abs/1983MNRAS.204..151L} {204, 151}

\bibitem[\protect\citeauthoryear{{Larson}}{{Larson}}{2010}]{larson10}
{Larson} R.~B.,  2010, \mn@doi [Nature Physics] {10.1038/nphys1484}, \href
  {http://adsabs.harvard.edu/abs/2010NatPh...6...96L} {6, 96}

\bibitem[\protect\citeauthoryear{{Lawrence}}{{Lawrence}}{1991}]{lawrence91}
{Lawrence} A.,  1991, \mn@doi [\mnras] {10.1093/mnras/252.4.586}, \href
  {http://adsabs.harvard.edu/abs/1991MNRAS.252..586L} {252, 586}

\bibitem[\protect\citeauthoryear{{Marin} \& {Antonucci}}{{Marin} \&
  {Antonucci}}{2016}]{ma16}
{Marin} F.,  {Antonucci} R.,  2016, \mn@doi [\apj]
  {10.3847/0004-637X/830/2/82}, \href
  {http://adsabs.harvard.edu/abs/2016ApJ...830...82M} {830, 82}

\bibitem[\protect\citeauthoryear{{Miley}}{{Miley}}{1968}]{miley68}
{Miley} G.~K.,  1968, \mn@doi [\nat] {10.1038/218933a0}, \href
  {http://adsabs.harvard.edu/abs/1968Natur.218..933M} {218, 933}

\bibitem[\protect\citeauthoryear{{Narayan} \& {Yi}}{{Narayan} \&
  {Yi}}{1994}]{ny94}
{Narayan} R.,  {Yi} I.,  1994, \mn@doi [\apjl] {10.1086/187381}, \href
  {http://adsabs.harvard.edu/abs/1994ApJ...428L..13N} {428, L13}

\bibitem[\protect\citeauthoryear{{Neeser}, {Eales}, {Law-Green}, {Leahy}  \&
  {Rawlings}}{{Neeser} et~al.}{1995}]{neeser95}
{Neeser} M.~J.,  {Eales} S.~A.,  {Law-Green} J.~D.,  {Leahy} J.~P.,
  {Rawlings} S.,  1995, \mn@doi [\apj] {10.1086/176201}, \href
  {http://adsabs.harvard.edu/abs/1995ApJ...451...76N} {451, 76}

\bibitem[\protect\citeauthoryear{{Ogle}, {Cohen}, {Miller}, {Tran}, {Fosbury}
  \& {Goodrich}}{{Ogle} et~al.}{1997}]{ogle97}
{Ogle} P.~M.,  {Cohen} M.~H.,  {Miller} J.~S.,  {Tran} H.~D.,  {Fosbury}
  R.~A.~E.,   {Goodrich} R.~W.,  1997, \mn@doi [\apjl] {10.1086/310675}, \href
  {http://adsabs.harvard.edu/abs/1997ApJ...482L..37O} {482, L37}

\bibitem[\protect\citeauthoryear{{Ogle}, {Whysong}  \& {Antonucci}}{{Ogle}
  et~al.}{2006}]{ogle06}
{Ogle} P.,  {Whysong} D.,   {Antonucci} R.,  2006, \mn@doi [\apj]
  {10.1086/505337}, \href {http://adsabs.harvard.edu/abs/2006ApJ...647..161O}
  {647, 161}

\bibitem[\protect\citeauthoryear{{Planck Collaboration} et~al.,}{{Planck
  Collaboration} et~al.}{2016}]{planckcosmo}
{Planck Collaboration} et~al., 2016, \mn@doi [\aap]
  {10.1051/0004-6361/201525830}, \href
  {http://adsabs.harvard.edu/abs/2016A%26A...594A..13P} {594, A13}

\bibitem[\protect\citeauthoryear{{Polletta} et~al.,}{{Polletta}
  et~al.}{2007}]{polletta07}
{Polletta} M.,  et~al., 2007, \mn@doi [\apj] {10.1086/518113}, \href
  {http://adsabs.harvard.edu/abs/2007ApJ...663...81P} {663, 81}

\bibitem[\protect\citeauthoryear{{Quataert}}{{Quataert}}{2001}]{quataert01}
{Quataert} E.,  2001, in {Peterson} B.~M.,  {Pogge} R.~W.,   {Polidan} R.~S.,
  eds,  Astronomical Society of the Pacific Conference Series Vol. 224, Probing
  the Physics of Active Galactic Nuclei. p.~71

\bibitem[\protect\citeauthoryear{{Rigby}, {Best}, {Brookes}, {Peacock},
  {Dunlop}, {R{\"o}ttgering}, {Wall}  \& {Ker}}{{Rigby} et~al.}{2011}]{rigby11}
{Rigby} E.~E.,  {Best} P.~N.,  {Brookes} M.~H.,  {Peacock} J.~A.,  {Dunlop}
  J.~S.,  {R{\"o}ttgering} H.~J.~A.,  {Wall} J.~V.,   {Ker} L.,  2011, \mn@doi
  [\mnras] {10.1111/j.1365-2966.2011.19167.x}, \href
  {http://adsabs.harvard.edu/abs/2011MNRAS.416.1900R} {416, 1900}

\bibitem[\protect\citeauthoryear{{Rigby}, {Argyle}, {Best}, {Rosario}  \&
  {R{\"o}ttgering}}{{Rigby} et~al.}{2015}]{rigby15}
{Rigby} E.~E.,  {Argyle} J.,  {Best} P.~N.,  {Rosario} D.,   {R{\"o}ttgering}
  H.~J.~A.,  2015, \mn@doi [\aap] {10.1051/0004-6361/201526475}, \href
  {http://adsabs.harvard.edu/abs/2015A%26A...581A..96R} {581, A96}

\bibitem[\protect\citeauthoryear{{Schmidt} \& {Smith}}{{Schmidt} \&
  {Smith}}{2000}]{schmidt00}
{Schmidt} G.~D.,  {Smith} P.~S.,  2000, \mn@doi [\apj] {10.1086/317811}, \href
  {http://adsabs.harvard.edu/abs/2000ApJ...545..117S} {545, 117}

\bibitem[\protect\citeauthoryear{{Shakura} \& {Sunyaev}}{{Shakura} \&
  {Sunyaev}}{1973}]{shakura73}
{Shakura} N.~I.,  {Sunyaev} R.~A.,  1973, \aap, \href
  {http://adsabs.harvard.edu/abs/1973A%26A....24..337S} {24, 337}
  
 \bibitem[\protect\citeauthoryear{{Shimwell} et~al.,}{{Shimwell}
  et~al.}{2017}]{lotss}
{Shimwell} T.~W.,  et~al., 2017, \mn@doi [\aap] {10.1051/0004-6361/201629313},
  \href {http://adsabs.harvard.edu/abs/2017A%26A...598A.104S} {598, A104}

\bibitem[\protect\citeauthoryear{{Singal}}{{Singal}}{2014}]{singal14}
{Singal} A.~K.,  2014, \mn@doi [\aj] {10.1088/0004-6256/148/1/16}, \href
  {http://adsabs.harvard.edu/abs/2014AJ....148...16S} {148, 16}

\bibitem[\protect\citeauthoryear{{Singal} \& {Singh}}{{Singal} \&
  {Singh}}{2013}]{singal13a}
{Singal} A.~K.,  {Singh} R.~L.,  2013, \mn@doi [\mnras]
  {10.1093/mnrasl/slt091}, \href
  {http://adsabs.harvard.edu/abs/2013MNRAS.435L..38S} {435, L38}

\bibitem[\protect\citeauthoryear{{Smith}}{{Smith}}{2015}]{weave}
{Smith} D.~J.~B.,  2015, preprint, \href
  {http://adsabs.harvard.edu/abs/2015arXiv150605630S} {} (\mn@eprint {arXiv}
  {1506.05630})

\bibitem[\protect\citeauthoryear{{Stern} et~al.,}{{Stern}
  et~al.}{2005}]{stern05}
{Stern} D.,  et~al., 2005, \mn@doi [\apj] {10.1086/432523}, \href
  {http://adsabs.harvard.edu/abs/2005ApJ...631..163S} {631, 163}

\bibitem[\protect\citeauthoryear{{Tasse}, {Best}, {R{\"o}ttgering}  \& {Le
  Borgne}}{{Tasse} et~al.}{2008}]{tasse08}
{Tasse} C.,  {Best} P.~N.,  {R{\"o}ttgering} H.,   {Le Borgne} D.,  2008,
  \mn@doi [\aap] {10.1051/0004-6361:20079299}, \href
  {http://adsabs.harvard.edu/abs/2008A%26A...490..893T} {490, 893}

\bibitem[\protect\citeauthoryear{{Urry} \& {Padovani}}{{Urry} \&
  {Padovani}}{1995}]{up95}
{Urry} C.~M.,  {Padovani} P.,  1995, \mn@doi [\pasp] {10.1086/133630}, \href
  {http://adsabs.harvard.edu/abs/1995PASP..107..803U} {107, 803}

\bibitem[\protect\citeauthoryear{{Wardle} \& {Miley}}{{Wardle} \&
  {Miley}}{1974}]{wm74}
{Wardle} J.~F.~C.,  {Miley} G.~K.,  1974, \aap, \href
  {http://adsabs.harvard.edu/abs/1974A%26A....30..305W} {30, 305}

\bibitem[\protect\citeauthoryear{{Whysong} \& {Antonucci}}{{Whysong} \&
  {Antonucci}}{2004}]{whysong04}
{Whysong} D.,  {Antonucci} R.,  2004, \mn@doi [\apj] {10.1086/380828}, \href
  {http://adsabs.harvard.edu/abs/2004ApJ...602..116W} {602, 116}

\bibitem[\protect\citeauthoryear{{Williams} et~al.,}{{Williams}
  et~al.}{2016}]{williams16}
{Williams} W.~L.,  et~al., 2016, \mn@doi [\mnras] {10.1093/mnras/stw1056},
  \href {http://adsabs.harvard.edu/abs/2016MNRAS.460.2385W} {460, 2385}

\bibitem[\protect\citeauthoryear{{Willott}, {Rawlings}, {Blundell}  \&
  {Lacy}}{{Willott} et~al.}{1999}]{willott99}
{Willott} C.~J.,  {Rawlings} S.,  {Blundell} K.~M.,   {Lacy} M.,  1999, \mn@doi
  [\mnras] {10.1046/j.1365-8711.1999.02907.x}, \href
  {http://adsabs.harvard.edu/abs/1999MNRAS.309.1017W} {309, 1017}

\bibitem[\protect\citeauthoryear{{Willott}, {Rawlings}, {Blundell}  \&
  {Lacy}}{{Willott} et~al.}{2000}]{willott00}
{Willott} C.~J.,  {Rawlings} S.,  {Blundell} K.~M.,   {Lacy} M.,  2000, \mn@doi
  [\mnras] {10.1046/j.1365-8711.2000.03447.x}, \href
  {http://adsabs.harvard.edu/abs/2000MNRAS.316..449W} {316, 449}

\bibitem[\protect\citeauthoryear{{Willott}, {Rawlings}, {Jarvis}  \&
  {Blundell}}{{Willott} et~al.}{2003}]{willott03}
{Willott} C.~J.,  {Rawlings} S.,  {Jarvis} M.~J.,   {Blundell} K.~M.,  2003,
  \mn@doi [\mnras] {10.1046/j.1365-8711.2003.06172.x}, \href
  {http://adsabs.harvard.edu/abs/2003MNRAS.339..173W} {339, 173}

\bibitem[\protect\citeauthoryear{{de Vries}, {Morganti}, {R{\"o}ttgering},
  {Vermeulen}, {van Breugel}, {Rengelink}  \& {Jarvis}}{{de Vries}
  et~al.}{2002}]{devries02}
{de Vries} W.~H.,  {Morganti} R.,  {R{\"o}ttgering} H.~J.~A.,  {Vermeulen} R.,
  {van Breugel} W.,  {Rengelink} R.,   {Jarvis} M.~J.,  2002, \mn@doi [\aj]
  {10.1086/338906}, \href {http://adsabs.harvard.edu/abs/2002AJ....123.1784D}
  {123, 1784}

\bibitem[\protect\citeauthoryear{{van Haarlem} et~al.,}{{van Haarlem}
  et~al.}{2013}]{vh13}
{van Haarlem} M.~P.,  et~al., 2013, \mn@doi [\aap]
  {10.1051/0004-6361/201220873}, \href
  {http://adsabs.harvard.edu/abs/2013A%26A...556A...2V} {556, A2}

\makeatother
\end{thebibliography}
\end{document}